\documentclass[prd,12pt]{article}
\pdfoutput=1

\usepackage{tikz}
\usepackage{amsmath,amssymb,graphicx,multirow,xspace,slashed,array,booktabs}
\usepackage[colorlinks=true,urlcolor=blue,anchorcolor=blue,citecolor=blue,filecolor=blue,linkcolor=blue,menucolor=blue,pagecolor=blue]{hyperref}
\usepackage[compress,numbers]{natbib}
\usepackage{subfigure, placeins}
\usepackage{booktabs}
\usepackage{blindtext, rotating}
\usepackage{afterpage}
\usepackage{enumitem}
\usepackage{ marvosym }
\usepackage{authblk} 
\usepackage{verbatim}
\usepackage{soul} %scratch out text
\usepackage[normalem]{ulem}
\usepackage{pifont}% http://ctan.org/pkg/pifont
\usepackage{booktabs}

\usepackage{floatrow}
\newfloatcommand{capbtabbox}{table}[][\FBwidth]

\usepackage[font=footnotesize,labelfont=bf]{caption}

\usepackage{lineno}

\allowdisplaybreaks

\long\def\symbolfootnote[#1]#2{\begingroup%
\def\thefootnote{\fnsymbol{footnote}}\footnote[#1]{#2}\endgroup}

\allowdisplaybreaks

\makeatletter
	
	\@addtoreset{equation}{section}
\makeatother

\newcommand{\newc}{\newcommand}
\newc{\gsim}{\lower.7ex\hbox{$\;\stackrel{\textstyle>}{\sim}\;$}}
\newc{\lsim}{\lower.7ex\hbox{$\;\stackrel{\textstyle<}{\sim}\;$}}
\newc{\gev}{\,{\rm GeV}}
\newc{\mev}{\,{\rm MeV}}
\newc{\ev}{\,{\rm eV}}
\newc{\kev}{\,{\rm keV}}
\newc{\tev}{\,{\rm TeV}}
\newc{\MHT}{$H_T^{\text{miss}}$}
\newc{\MET}{$\slashed{E}_T$}
\newc{\MTT}{$M_{T2}$}

\newc{\mz}{M_Z}
\newc{\mpl}{M_*}
\newc{\mw}{m_{\rm weak}}
\newc{\nr}[1]{N^c_R{}_{#1}}

\def\beq{\begin{equation}}
\def\eeq{\end{equation}}
\newcommand{\bea}{\begin{eqnarray}\begin{aligned}}
\newcommand{\eea}{\end{aligned}\end{eqnarray}}
\def\bitem{\begin{itemize}}
\def\eitem{\end{itemize}}

\begin{document}
\baselineskip 0.6cm

\begin{titlepage}

\vspace*{-0.5cm}

\thispagestyle{empty}

\begin{center}

\vskip 1cm

{\LARGE \bf
Multiple Hierarchies \\[1ex] from a Warped Extra Dimension
}

\vskip 1cm

\vskip 1.0cm
{\large Seung J. Lee$^1$, Yuichiro Nakai$^2$ and Motoo Suzuki$^2$}
\vskip 1.0cm
{\it
$^1$Department of Physics, Korea University, Seoul 136-713, Korea\\
$^2$Tsung-Dao Lee Institute and School of Physics and Astronomy, \\Shanghai
Jiao Tong University, 800 Dongchuan Road, Shanghai, 200240 China \\}
\vskip 1.0cm

\end{center}

\vskip 1cm

\begin{abstract}

Theories beyond the Standard Model often contain mass scales hierarchically different from
the electroweak scale and the Planck scale. 
It has been shown that such
hierarchical mass scales 
can be
realized as typical energy scales of multiple 3-branes in a 5D warped spacetime.
We present a mechanism for stabilizing the intervals between the multiple 3-branes in the warped extra dimension,
by introducing a single 5D scalar field with brane-localized potentials.
We discuss the radion stabilization by solving the Einstein equation and the scalar field equation of motion
so that a backreaction effect on the geometry due to the presence of the scalar field is taken into account.
Perturbations from the background configuration are then considered with proper identification of multiple radion degrees of freedom.
By solving their equations of motion, we compute the mass spectrum of the radion-scalar field system
and the radion couplings to brane-localized matter fields,
which are found to be suppressed by typical energy scales and radion profiles at the branes.
We also compute the mass spectrum of Kaluza-Klein gravitons and their profiles in the extra dimension. 
Some applications of the setup are briefly described.
Our analysis provides a solid ground to build 5D warped extra dimension models with multiple 3-branes.

\end{abstract}

\flushbottom

\end{titlepage}

%#######################
\section{Introduction}\label{intro}

Laws of nature spread over a wide range of energy scales.
There is a huge hierarchy between the Planck scale and the electroweak scale
whose stability under radiative corrections is problematic in the Standard Model (SM). 
Moreover, the existence of new mass scales hierarchically different from the Planck scale and the electroweak scale
is often indicated by theories of physics beyond the SM.
For instance, the (invisible) axion solution to the strong CP problem
\cite{Peccei:1977hh,Weinberg:1977ma,Wilczek:1977pj,Kim:1979if,Shifman:1979if,Dine:1981rt,Zhitnitsky:1980tq}
introduces an energy scale
where the Peccei-Quinn (PQ) symmetry is spontaneously broken.
If supersymmetry (SUSY) is realized in nature, it must be broken at an intermediate scale
\cite{Martin:1997ns}.
In addition, dark matter may indicate the existence of a new mass scale hierarchically smaller than the electroweak scale
\cite{Battaglieri:2017aum}.

A large hierarchy between two energy scales is naturally generated in the Randall-Sundrum (RS) model \cite{Randall:1999ee}
whose geometry consists of $\rm AdS_5$ bulk spacetime bounded by two 3-branes with positive and negative tensions
called the UV and IR branes, respectively.
Due to a warp factor, the energy scale of the IR brane is exponentially smaller than that of the UV brane.
In the original RS model, the distance between the two branes is a free parameter,
and the corresponding modulus field called {\it radion} is massless.
The radion stabilization has been explored in
refs.~\cite{Goldberger:1999uk,Garriga:2000jb,Goldberger:2000dv,Hofmann:2000cj,Brevik:2000vt,Luty:2000ec,Flachi:2001pq,Nojiri:2001ai,Garriga:2002vf,Haba:2019zjc,Fujikura:2019oyi},
among which the Goldberger-Wise (GW) mechanism \cite{Goldberger:1999uk} is the most popular.
The GW mechanism introduces a bulk scalar field with potentials localized on the UV and IR branes.
The potentials make the scalar field develop nonzero vacuum expectation values (VEVs) on the branes.
The GW scalar field then obtains a nontrivial bulk profile so that a radion potential is generated.
Properties of the radion field in the GW mechanism have been investigated in~\cite{Csaki:1999mp,Goldberger:1999un}
by using a naive ansatz which ignores the radion wavefunction and the backreaction of the GW field on the geometry,
and a more rigorous treatment of these effects has been presented in~\cite{Csaki:2000zn}.
The AdS/CFT correspondence
\cite{Maldacena:1997re,Gubser:1998bc,Witten:1998qj} tells us that
the RS model can be understood by a nearly-conformal strongly-coupled 4D field theory
\cite{ArkaniHamed:2000ds,Rattazzi:2000hs}.
How the presence of couplings of the bulk SM fields to the GW scalar modifies the identification of the parameters on the two sides of the AdS/CFT correspondence has been presented in ref.~\cite{Chacko:2014pqa}.
The introduction of the GW scalar field corresponds to a small relevant deformation of the 4D CFT
which triggers spontaneous breaking of the conformal symmetry at the IR scale. 

The RS model is generalized by introducing more than two 3-branes with positive or negative tensions in a 5D warped spacetime
to realize multiple hierarchical mass scales.
The energy scale of each brane is exponentially different from those of the other branes.
Depending on the size of a bulk cosmological constant in each subregion bounded by two branes,
the warp factor of each subregion can take a different value.
Corresponding to each interval between two branes, a radion degree of freedom exists. 
Such multi-brane setups have been considered in~\cite{Lykken:1999nb,Hatanaka:1999ac,Kogan:1999wc,Gregory:2000jc,Kogan:2000cv,Mouslopoulos:2000er,Kogan:2000xc},
originally motivated by cosmology in warped extra dimension models.
Refs.~\cite{Mouslopoulos:2001uc,Kogan:2001wp} have studied the wavefunction and mass spectrum of a bulk matter field.
One can also find applications to the flavor structure of the SM~\cite{Oda:1999di,Oda:1999be,Dvali:2000ha,Moreau:2004qe}
and collider phenomenology~\cite{Agashe:2016rle,Agashe:2016kfr,Csaki:2016kqr}.
Further applications of the multi-brane setups are likely to be considered, and
there is a huge potential for innovation in model-building of physics beyond the SM.
However, thorough studies of multi-brane models with stabilized radions have not been conducted so far. 
Refs.~\cite{Pilo:2000et,Kogan:2001qx} initiated to analyze the dynamics of radions in a three 3-brane system
without considering their stabilization. 
Ref.~\cite{Choudhury:2000wc} applied the GW mechanism to the stabilization of radions in a three 3-brane system
different from what we consider here in terms of bulk cosmological constants and signs of brane tensions,
and they used the naive ansatz ignoring the radion wavefunctions and the backreaction of the GW field on the geometry.
Therefore, a more complete investigation of multi-brane models is necessary.

In this paper, we discuss the stabilization of radions in multi-brane models
through a simple extension of the GW mechanism with introducing a single 5D scalar field.
Each bulk subregion bounded by two branes has a different cosmological constant in general
which leads to a different warp factor.
For the sake of simplicity, our main focus is on a 5D model with three 3-branes whose tensions are
positive, positive and negative.
They are identified as the UV, intermediate and IR branes, respectively.
Generalization to any number of branes with positive or negative tensions is straightforward
and briefly described.
The Einstein equation and the scalar field equation of motion are simultaneously solved
so that the radion wavefunctions and the backreaction effect on the geometry are taken into account.
We then consider perturbations from the background configuration and
compute the mass spectrum of the radion-scalar field system by solving the equations of motion.
The radion couplings to brane-localized matter fields are also presented. 
Moreover, we compute the mass spectrum of Kaluza-Klein (KK) gravitons and their profiles in the extra dimension.
In the dual CFT picture of the three 3-brane model, the presence of the intermediate brane can be understood
as spontaneous breaking of a conformal symmetry, but the resulting 4D dual theory flows into a new conformal fixed point
\cite{Agashe:2016rle}.
The conformal symmetry is spontaneously broken again at the IR scale.

The rest of the paper is organized as follows.
In section~\ref{sec:background_metric}, the setup is introduced and
the radion stabilization with a single 5D scalar field is presented.
Our discussion focuses on the three 3-brane model, but we also comment on generalization to an arbitrary number of branes.
In section~\ref{sec:metric_perturbation}, perturbations from the background configuration are considered and
their equations of motion are presented.
Then, in section~\ref{sec:KK_towers}, we compute the KK mass spectrum of the radion-scalar field system
and also the spectrum of KK gravitons by solving the equations of motion.
Section~\ref{sec:radion_mass} finds nonzero masses of two radions in the three 3-brane model
by taking account of the backreaction effect of the GW scalar field background.
In section~\ref{sec:radioneffaction}, we present the radion effective action and
the radion couplings to brane-localized matter fields.
Section~\ref{discussions} is devoted to conclusions and discussions.
A few applications of the three 3-brane model are briefly described.
Appendix~\ref{app:naive_ansatz} presents the radion stabilization for the three 3-brane model by using the naive ansatz.
We consider a general case that each bulk subregion bounded by two branes has a different cosmological constant.
Some other details are discussed in appendices~\ref{app:generalN} and \ref{app:hermiticity}.

%#######################
\section{Background configuration}
\label{sec:background_metric}

\begin{figure}[!t]
    \includegraphics[width=6in]{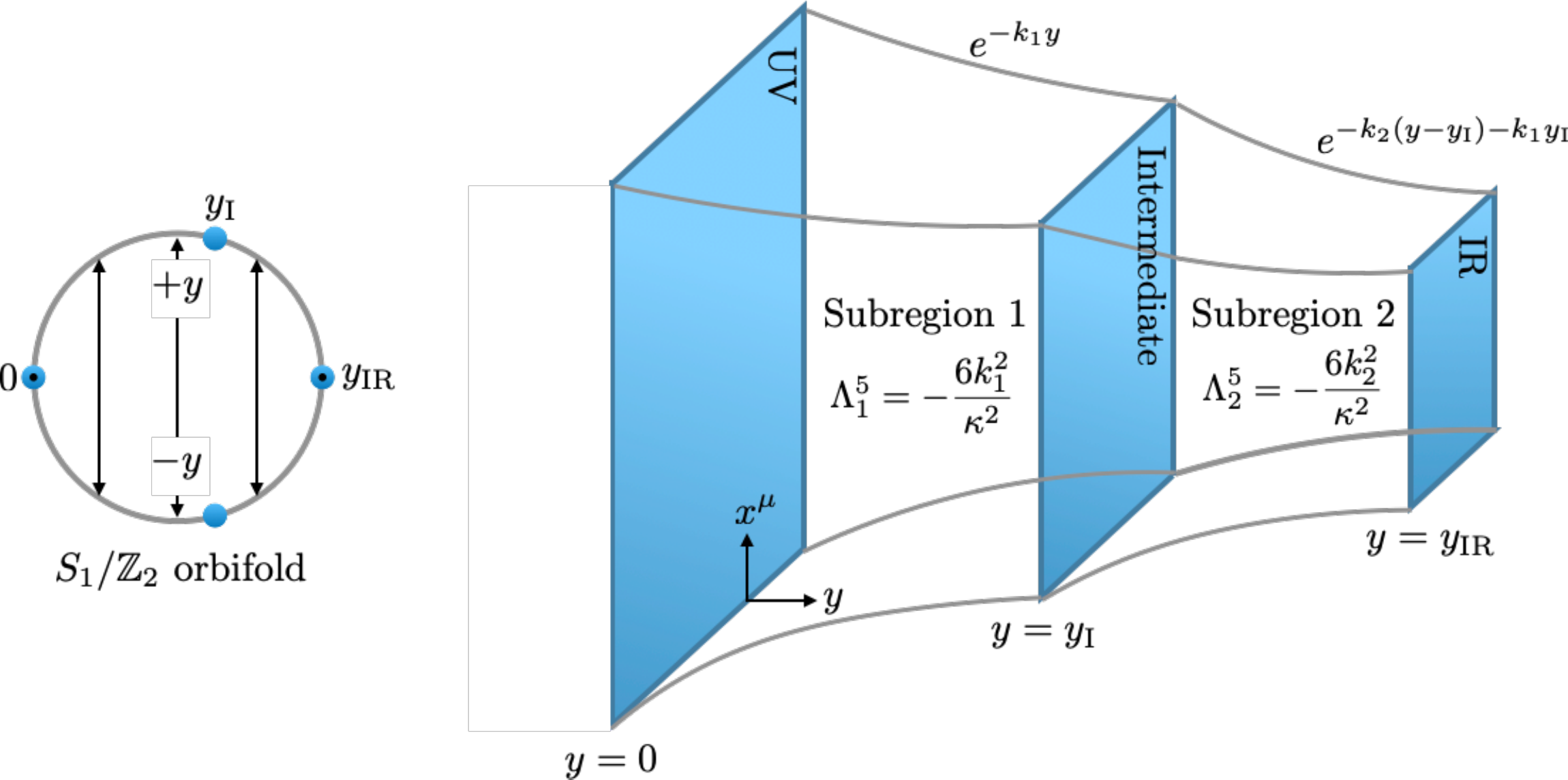}
    \vspace{0.5cm}
  \caption{Schematic pictures of the three 3-brane system.
  The left panel denotes the $S_1/ \mathbb{Z}_2$ orbifold extra dimension where $+y$ and $-y$ are identified
  and $y=\{0, y_{\rm IR}\}$ are the fixed points.
  The UV, intermediate and IR branes sit on $y=\{0,y_{\rm I},y_{\rm IR}\}$, respectively.
  In the right panel, we refer to the bulk region between the UV and intermediate (intermediate and IR) branes
  as the subregion $1$ (subregion $2$).
  The bulk cosmological constants $\Lambda_{1,2}^2$ and the warp factors,
  $e^{-k_1 y}$ and $e^{-k_2 (y-y_{\rm I})-k_1y_{\rm I}}$, in the subregions $1,2$ are different in general.}
  \label{fig:three_branes}
\end{figure}

We consider a spacetime geometry described by $\mathbb{R}^4 \times S_1/ \mathbb{Z}_2$ with the metric,
\begin{align}
\label{metric}
ds^2= g_{MN} dx^M dx^N = e^{-2A(y)}\eta_{\mu\nu}dx^\mu dx^\nu-dy^2\ .
\end{align}
Here, $M=(\mu, y)$ with $\mu$ running from 0 to 3, $\eta_{\mu\nu}$ is the flat 4D metric,
$y \in [0,\,y_{\rm IR}]$ denotes the coordinate
for the $S_1/ \mathbb{Z}_2$ orbifold and $A(y)$ is some function of $y$.
The left panel of Fig.~\ref{fig:three_branes} shows a schematic picture of the orbifold extra dimension.
Two 3-branes extending over $\mathbb{R}^4$, which we call UV and IR branes,
reside on the orbifold fixed points, $y = y_{\rm UV} = 0$ and $y= y_{\rm IR}$, respectively.
In addition, we introduce another intermediate 3-brane placed at a point $y=y_{\rm I}$ between the boundaries.
We would like to stabilize the distance between the UV and intermediate branes $y_{\rm I}$ 
and the distance between the intermediate and IR branes $y_{\rm IR} - y_{\rm I}$.
As in the case of the GW mechanism for the two 3-brane model, 
a bulk real scalar field plays that role.
The action with the scalar field labeled by $\phi$ (the GW field) is given by
\begin{equation}
\begin{split}
\label{eq:action}
S=&-2M_5^3\int d^4 x dy\sqrt{g} \, R
+\int d^4 x dy\sqrt{g}\left(\frac{1}{2}\nabla\phi\nabla\phi-V(\phi)\right)\\[1ex]
&-\int d^4 x\sqrt{|g^{in}_{\rm UV}|}\,\chi_{\rm UV}
-\int d^4 x\sqrt{|g^{in}_{\rm I}|}\,\chi_{\rm I}
-\int d^4 x\sqrt{|g^{in}_{\rm IR}|}\,\chi_{\rm IR}\ ,
\end{split}
\end{equation}
where $M_5$ is the five-dimensional Planck mass, $R = g^{MN}R_{MN}$ is the Ricci scalar,
$\sqrt{g}$ and $\sqrt{|g^{in}|}$ are the volume elements of the bulk metric and
the induced metric on each brane, respectively, and $\chi$'s are brane-localized potentials.
The subscripts $\rm UV,I,IR$ indicate the UV, intermediate and IR branes, respectively.
$V(\phi)$ denotes the potential of the scalar field
which includes bulk cosmological constants.
Following the superpotential method discussed in ref.~\cite{DeWolfe:1999cp},
we assume the potential takes a specific form,
\begin{align}
\label{scalarpotential}
V(\phi)&=\frac{1}{8}\left(\frac{\partial W}{\partial\phi}\right)^2-\frac{\kappa^2}{6}W(\phi)^2\ ,
\end{align}
where $\kappa^2=(4M^3_5)^{-1}$ and
$W(\phi)$ is given by different forms for the bulk region between the UV and intermediate branes (subregion 1)
and for the region between the intermediate and IR branes (subregion 2),
\begin{align}
\label{Wphi}
W(\phi)=\left\{
\begin{array}{c}
\frac{6k_1}{\kappa^2}-u_1\phi^2 \quad (\text{subregion~$1$}) \\[1ex]
\frac{6k_2}{\kappa^2}-u_2\phi^2 \quad (\text{subregion~$2$})
\end{array}
\right.  .
\end{align}
Here, $k_{1,2}$ and $u_{1,2}$ are positive constants with mass dimension 1.
Inserting these expressions into Eq.~\eqref{scalarpotential},
the bulk cosmological constants for the subregions $1,2$ are given by
\begin{align}
\Lambda_1^5=-\frac{6k_1^2}{\kappa^2} \quad (\text{subregion~$1$}) \, , \qquad
\Lambda_2^5=-\frac{6k_2^2}{\kappa^2} \quad (\text{subregion~$2$}) \, .
\end{align}
Note that the two bulk cosmological constants $\Lambda_{1,2}$ are different in general.
The brane-localized potentials are taken as
\begin{align}
&\chi_{\rm UV}(\phi)=W(\phi_{\rm UV})+\frac{\partial W(\phi_{\rm UV})}{\partial\phi}(\phi-\phi_{\rm UV})+\gamma^2_{\rm UV}(\phi-\phi_{\rm UV})^2\ , \nonumber\\[1ex]
&\chi_{\rm I}(\phi)=
\frac{1}{2}\left[W(\phi_{\rm I+})-W(\phi_{\rm I-})\right]
+\frac{1}{2}\left[\frac{\partial W(\phi_{\rm I+})}{\partial\phi}
-\frac{\partial W(\phi_{\rm I-})}{\partial\phi}\right](\phi-\phi_{\rm I})
+\gamma^2_{\rm I}(\phi-\phi_{\rm I})^2\ , \nonumber \\[1ex]
&\chi_{\rm IR}(\phi)=-W(\phi_{\rm IR})-\frac{\partial W(\phi_{\rm IR})}{\partial\phi}(\phi-\phi_{\rm IR})+\gamma^2_{\rm IR}(\phi-\phi_{\rm IR})^2\ ,
\label{branelocalizedpotential}
\end{align}
where all the $\gamma^2$'s are constants with mass dimension $1$
and $\phi_{\rm UV, \,I, \,IR} $ are values of $\phi$ at the UV, intermediate and IR branes, respectively. 
For the intermediate brane, the subscripts $\rm I+$ and $\rm I-$ indicate values
in the limit of $y=\lim_{\epsilon\to 0+} [y_{\rm I}-\epsilon]$ and $y=\lim_{\epsilon\to 0-} [y_{\rm I}-\epsilon]$.
The setup is summarized in the right panel of Fig.~\ref{fig:three_branes}.

Let us now find a background configuration for $A(y)$ and $\phi(x,y)$.
The scalar field respects the 4D Lorentz invariance,
\begin{align}
\label{phi0y}
\phi(x,y) =\phi_0(y) .
\end{align}
Then, the Einstein equation,
\begin{align}
    R_{MN}=\kappa^2\left(T_{MN}-\frac{1}{3}g_{MN}T\right) ,
\end{align}
with the trace of the energy-momentum tensor $T=g^{KL}T_{KL}$
and the field equation for $\phi$ lead to coupled equations,
\begin{equation}
\begin{split}
\label{coupledequations}
4A'^2-A''&=-\frac{2}{3}\kappa^2 V(\phi_0)-\frac{\kappa^2}{3}\sum_i \chi_i(\phi_0)\,\delta(y-y_i)\ ,\\
A'^2&=\frac{\kappa^2\phi'^2_0}{12}-\frac{\kappa^2}{6}V(\phi_0)\ ,\\[1ex]
\phi''_0&=4A'\phi_0'+\frac{\partial V(\phi_0)}{\partial\phi_0}+\sum_i\frac{\partial\chi_i(\phi_0)}{\partial\phi_0}\,\delta(y-y_i)\ ,
\end{split}
\end{equation}
where the prime $'$ denotes the partial derivative with respect to $y$ $i.e.$ $\partial/\partial y$
and $i= \rm UV,I,IR$.
In the above equations, we match the singular terms on each brane by imposing the following boundary conditions:  
\begin{align}
\label{eq:bound_Ap}
    \left[A'\right]|_{y=y_i}=\frac{\kappa^2}{3}\chi_i(\phi_0)\ , \qquad 
\left[\phi_0'\right]|_{y=y_i}=\frac{\partial\chi_i(\phi_0)}{\partial\phi_0}\ .
\end{align}
Here, we have defined $\left[X\right]|_{y=y_i}\equiv \lim_{\epsilon\to+0}\left[ X(y_i+\epsilon)- X(y_i-\epsilon)\right]$. 
With the potential form of Eq.~\eqref{scalarpotential}, a solution to the first-order differential equations,
\begin{align}
  \phi'_0=\frac{1}{2}\frac{\partial W(\phi_0)}{\partial\phi_0}\ , \qquad  A'=\frac{\kappa^2}{6}W(\phi_0)\ ,
\end{align}
gives a solution to the coupled equations \eqref{coupledequations}
satisfying the boundary conditions \eqref{eq:bound_Ap}.
By using the explicit forms of $W(\phi)$ in Eq.~\eqref{Wphi} and the brane-localized potentials \eqref{branelocalizedpotential},
the solution is obtained as
\begin{equation}
\begin{split}
\label{backgroundsolutions}
\phi_0 (y)&=\left\{
\begin{array}{c}
\phi_{\rm UV}\,e^{-u_1 y} \quad (\text{subregion~$1$})\\
\phi_{\rm UV} \,e^{-u_2y+y_{\rm I}(u_2-u_1)} \quad (\text{subregion~$2$}) \ ,
\end{array}
\right.\\[1ex]
A (y)&=\left\{
\begin{array}{c}
k_1 y+\frac{\kappa^2\phi_{\rm UV}^2}{12}e^{-2u_1y} \quad (\text{subregion~$1$})\\
k_2 y+y_{\rm I}(k_1-k_2)+\frac{\kappa^2\phi_{\rm UV}^2}{12}e^{-2u_2y+2y_{\rm I}(u_2-u_1)} \quad (\text{subregion~$2$})\ .
\end{array}
\right. 
\end{split}
\end{equation}
We can see that the first term in the solution of $A(y)$ for each subregion gives a warp factor
(see also the right panel of Fig.~\ref{fig:three_branes}).
The last term indicates a backreaction effect from the GW field.
It should be also noted that, if we take 
$k_2=k_1$ and ignore the terms including the GW field,
the intermediate brane disappears and
the RS geometry of the two 3-brane model is obtained. 
From the solution of $\phi_0 (y)$ for the subregion 1,
the distance between the UV and intermediate branes is related to the values of $\phi$ on the branes,
\begin{align}
\label{eq:GW_y_I}
   y_{\rm I}= \frac{1}{u_1}\log\left[\phi_{\rm UV}/\phi_{\rm I}\right] .
\end{align}
Similarly, the solution of $\phi_0 (y)$ for the subregion 2 gives the distance between the intermediate and IR branes,
\begin{align}
\label{eq:GW_y_IR}
   y_{\rm IR}-y_{\rm I}= \frac{1}{u_2}\log\left[\phi_{\rm I}/\phi_{\rm IR}\right] .
\end{align}
Therefore, all the distances between the 3-branes are determined in terms of the values of $\phi$ on the branes
by solving the Einstein equation and the scalar field equation of motion simultaneously.
The 4D Planck scale is now given by
\begin{align}
M_{\rm pl}^2 = \frac{M_5^3}{k_1}\left[1-e^{-2k_1y_{\rm I}}\left(\frac{1}{k_1}-\frac{1}{k_2}\right)
-\frac{e^{-2k_2(y_{\rm IR}-y_{\rm I})-2k_1y_{\rm I}}}{k_2}\right]   .
\end{align}
In appendix~\ref{app:naive_ansatz}, we present the determination of the brane separations by following the similar way as
the original discussion of the GW mechanism~\cite{Goldberger:1999uk,Goldberger:1999un} using the naive ansatz for the metric in an effective theory approach.

\begin{figure}[!t]
    \includegraphics[width=5.5in]{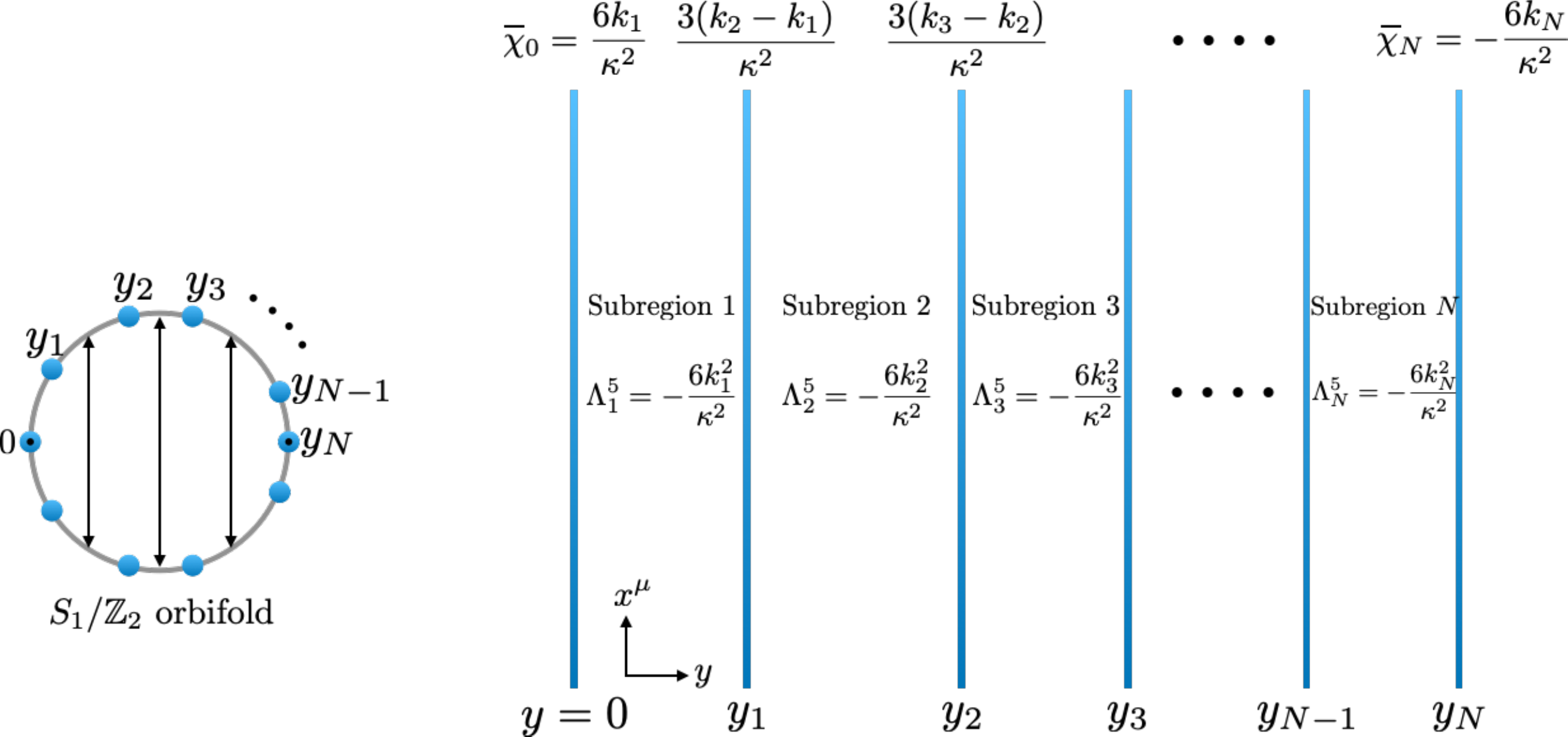}
    \vspace{0.5cm}
  \caption{Schematic pictures of the $N+1$ 3-brane system.
  The left panel denotes the $S_1/ \mathbb{Z}_2$ orbifold extra dimension where $+y$ and $-y$ are identified
  and $y=0, y_N$ are the fixed points.
  The $N+1$ 3-branes are located at $y=0,y_1,y_2,\cdots,y_{N-1},y_N$.
  In the right panel, $\Lambda^5_p$ denotes the bulk cosmological constant in the subregion $p$ defined in $y_{p-1}< y < y_p$.
  The brane tension $\bar\chi_p$ for the brane at $y=y_p$ is obtained
  by ignoring the effect of the GW field on the brane-localized potential.
  }
  \label{fig:N_1_branes}
\end{figure}

The discussion can be generalized by placing $N+1$ 3-branes at points $y = y_0\,(=0),y_1,y_2, \cdots,y_{N-1},y_N$
where $y_0,y_N$ are the orbifold fixed points (see the left panel of Fig.~\ref{fig:N_1_branes}).
The potential of the GW scalar field including bulk cosmological constants
is assumed to have the same form as Eq.~\eqref{scalarpotential}
with $W(\phi)$ for the subregion $p$ defined in $y_{p-1} < y < y_p$,
\begin{align}
    W_p(\phi)=\frac{6k_p}{\kappa^2}-u_p \phi^2 \quad (\text{subregion~$p$}) \ ,
\end{align}
where $k_p, u_p$ are positive constants with mass dimension 1. 
The bulk cosmological constant for the subregion $p$ is then given by
\begin{align}
    \Lambda^5_p=-\frac{6k_p^2}{\kappa^2} \quad (\text{subregion~$p$}) \ .
\end{align}
The brane-localized potentials are also generalized as
\begin{align}
&\chi_{0}(\phi)=W_1(\phi_{0,1}(0))+\frac{\partial W_1(\phi_{0,1}(0))}{\partial\phi}(\phi-\phi_{0,1}(0))+\gamma^2_{0}(\phi-\phi_{0,1}(0))^2\ , \nonumber \\[1ex]
&\qquad \qquad \qquad \cdots \nonumber \\[1ex]
&\chi_{p}(\phi)=
\frac{1}{2}\left[W_{p+1}(\phi_{0,p+1}(y_p))-W_p(\phi_{0,p}(y_p))\right]  \\[1ex]
&~~~~~~\quad +\frac{1}{2}\left[\frac{\partial W_{p+1}(\phi_{0,p+1}(y_p))}{\partial\phi}
-\frac{\partial W_p(\phi_{0,p}(y_p))}{\partial\phi}\right](\phi-\phi_{0,p}(y_p))
+\gamma^2_{p}(\phi-\phi_{0,p}(y_p))^2\ ,  \nonumber \\[1ex]
&\qquad \qquad \qquad \cdots \nonumber \\[1ex]
&\chi_{N}(\phi)=-W_N(\phi_{0,N}(y_N))-\frac{\partial W_N(\phi_{0,N}(y_N))}{\partial\phi}(\phi-\phi_{0,N}(y_N))+\gamma^2_N(\phi-\phi_{0,N}(y_N))^2\ . \nonumber 
\end{align}
Here, $e.g.$ $\phi_{0,p}(y_p)$ denotes the value of $\phi$ on the brane placed at $y=y_p$.
The setup is illustrated schematically in the right panel of Fig.~\ref{fig:N_1_branes}.
Solving the Einstein equation and the scalar field equation of motion,
$\phi_0 (y)$ and $A(y)$ for the subregion $p$ are given by
\begin{equation}
\begin{split}
   & \phi_{0,p} (y) =\phi_{0,1}(0)e^{-u_p y+\sum_{q=1}^{p}(u_q-u_{q-1})y_{q-1}} ,\\[1ex]
   & A_p (y) =-k_p y+\sum_{q=1}^{p}(k_q-k_{q-1})y_{q-1}+\frac{\kappa^2}{12}\phi_{0,p} (y)^2\ .
\end{split}
\end{equation}
As in the case of the three 3-brane model,
the first two terms in the solution of $A_p(y)$ give a warp factor
while the last term indicates a backreaction effect from the GW field. 
The distance between the branes at $y_{p-1}$ and $y_p$ is determined as
\begin{align}
    y_p-y_{p-1}=\frac{1}{u_p}\log\left[\phi_{0,p}(y_{p-1})/\phi_{0,p}(y_{p})\right] \ ,
\end{align}
in terms of the scalar field values on the branes $\phi_{0,p}(y_{p-1}),\phi_{0,p}(y_{p})$.

%#######################
\section{Perturbations about the background}
\label{sec:metric_perturbation}
We now discuss perturbations about the background configuration obtained in the previous section.
Our focus here is on spin-$0$ fluctuations relevant for the stabilization of the inter-brane separations.
We derive the (linearized) equations of motion and boundary conditions that the fluctuations obey.
The similar discussion in the two 3-brane system with a GW field was presented in ref.~\cite{Csaki:2000zn}.
We extend the method to discuss the three 3-brane system.

The spin-$0$ fluctuations are introduced by taking a general ansatz,
\begin{equation}
\begin{split}
\label{eq:metric}
&\phi=\phi_0(y)+\hat\phi(x,y)\ ,\\[1ex]
&ds^2=e^{-2A(y)}\left[(1-2F(x,y))\eta_{\mu\nu}+E(x,y)_{,\mu\nu}\right]dx^\mu dx^\nu-\left[1+2G(x,y)\right]dy^2\ .
\end{split}
\end{equation}
Here, $\hat\phi(x,y),~F(x,y),~E(x,y),~G(x,y)$ denote the fluctuations, and $E_{,\mu\nu}=\partial_\mu\partial_\nu E$.
We note that one of the fluctuations $E(x,y)$ is trivial in the two 3-brane system and
it can be taken to zero over the whole spacetime (see $e.g.$ ref.~\cite{Charmousis:1999rg,Csaki:2000zn}).
In the three 3-brane model, on the other hand, a non-trivial distribution of $E(x,y)$ is required
to obtain non-trivial solutions to the equations that the perturbations obey, as we will see later.
By using Eq.~\eqref{eq:metric} and defining $\widetilde T_{MN}=T_{MN}-\frac{1}{3}Tg_{MN}$,
the linearized Einstein equation for the fluctuations is given by
\begin{align}
\label{linearized_Einstein_equation}
    \delta R_{MN}=\kappa^2 \delta \widetilde T_{MN}\ ,
\end{align}
where
\begin{align}
\label{deltaRT}
\delta R_{\mu\nu}&=\eta_{\mu\nu}\Box F+\eta_{\mu\nu}e^{-2A}\biggl[-F''+A''(2F+2G)+A'(8F'+G') \nonumber\\
&\qquad \qquad \qquad \qquad \qquad \quad \quad +A'^2(-8G-8F)-\frac{1}{2}A'\Box E' \biggr]\nonumber\\
&\quad +2F_{,\mu\nu}-G_{,\mu\nu}+e^{-2A}\left[\frac{1}{2}E''_{,\mu\nu}-A'' E_{,\mu\nu}+4A'^2 E_{,\mu\nu}-2A' E'_{,\mu\nu}\right]\ ,
\nonumber \\[1ex]
\delta R_{\mu 5}&=3 F'_{,\mu}-3A' G_{,\mu}\ , \nonumber  \\[1ex]
\delta R_{55}&=e^{2A}\Box G+4F''+A'(-8F'-4G')-\frac{1}{2}\Box E''+A'\Box E'\ , \\[1ex]
\begin{split}
\delta \widetilde T_{\mu\nu}&=-\frac{1}{3}e^{-2A}\sum_i\delta(y-y_i)\left[\eta_{\mu\nu}\left(\frac{\partial\chi_i}{\partial\phi}\hat\phi-2F\chi_i-G\chi_i\right)+E_{,\mu\nu}\chi_i\right]\\
&\quad -\frac{2}{3}e^{-2A}\left[\eta_{\mu\nu}\left(\frac{\partial V}{\partial\phi}\hat\phi-2F V\right)+E_{,\mu\nu}V\right]\ ,
\end{split}
\nonumber \\[2ex]
\delta\widetilde T_{55}&=2\phi_0'\hat\phi'+\frac{2}{3}\left[2GV+\frac{\partial V}{\partial\phi}\hat\phi\right]+\frac{4}{3}\sum_i\delta(y-y_i)\left[G\chi_i+\frac{\partial\chi_i}{\partial\phi}\hat\phi\right]\ , \nonumber  \\[1ex]
\delta \widetilde T_{\mu5}&=\partial_\mu\hat\phi\,\phi_0'\ , \nonumber
\end{align}
with $i= \rm UV,I,IR$ and $\Box \equiv \eta^{\mu\nu}\partial_\mu\partial_\nu $.
The linearized field equation for $\hat\phi$ gives
\begin{equation}
\begin{split}
    -\hat\phi''+\,&e^{2A}\Box\hat\phi+4A'\hat\phi'+\frac{\partial^2 V}{\partial\phi^2}\hat\phi \\
    &=-\sum_i\delta(y-y_i)\left[\frac{\partial^2\chi_i}{\partial\phi^2}\hat\phi+G\frac{\partial\chi_i}{\partial\phi}\right]-2G\frac{\partial V}{\partial\phi}-(G'+4F'-\frac{1}{2}\Box E')\phi_0'\ .
\label{linearized_field_equation}
\end{split}
\end{equation}
They are the coupled field equations we will solve. 

The boundary conditions that the perturbations obey on the branes are obtained
by matching the delta functions
in the Einstein equation~\eqref{linearized_Einstein_equation} with Eq.~\eqref{deltaRT} and
the field equation~\eqref{linearized_field_equation},
\begin{align}
&[F']|_{y=y_i}-\frac{1}{8}[\Box E']|_{y=y_i}=\frac{1}{3}\kappa^2\left(G\chi_i+\frac{\partial\chi_i}{\partial\phi}\hat\phi\right) , \label{Eprimeboundary} \\[1ex]
&[E']|_{y=y_i}=0\ ,
 \label{Eprimemunuboundary} \\[1ex]
&[F']|_{y=y_i}=\frac{\kappa^2}{3}\chi_i G+\frac{\kappa^2}{3}\frac{\partial \chi_i}{\partial\phi}\hat\phi\ , \label{Fprimeboundary} \\[1ex]
&[\hat\phi']|_{y=y_i}=\frac{\partial^2\chi_i}{\partial\phi^2}\hat\phi+\frac{\partial\chi_i}{\partial\phi}G\ .
\label{phiprimeboundary}
\end{align}
The last condition \eqref{phiprimeboundary} is simplified by considering steep brane-localized potentials $\chi_i$
as in the case of the two 3-brane system~\cite{Csaki:2000zn}.
In the limit of $\partial^2\chi_i/\partial \phi^2\gg 1$, Eq.~\eqref{phiprimeboundary} is reduced to 
$\hat \phi|_{y=y_i}=0$.
We will use this condition in the following discussion. 
From the linearized Einstein equation $\delta R_{\mu 5}=\kappa^2\delta \widetilde{T}_{\mu 5}$, we require
$3F'-3A' G=\hat \phi \phi'$
which leads to the conditions on the branes,
$(F'-A' G)|_{y=y_i}=0$.
Then, by using Eq.~\eqref{eq:bound_Ap},
the boundary condition \eqref{Fprimeboundary} is automatically satisfied.
The boundary condition \eqref{Eprimeboundary} is reduced to \eqref{Fprimeboundary} by using Eq.~\eqref{Eprimemunuboundary}.
Finally, the condition \eqref{Eprimemunuboundary} leads to $E'|_{y=y_{\rm UV},y_{\rm IR}}=0$.
Then, the boundary conditions for the perturbations are summarized as
\begin{align}
\label{eq:b_phi}
&\hat\phi|_{y=y_{\rm UV},y_{\rm I},y_{\rm IR}}=0\ ,\\[1ex]
\label{eq:b_fg}
&F'-GA'|_{y=y_{\rm UV},y_{\rm I},y_{\rm IR}}=0\ ,\\[1ex]
\label{eq:b_ep_f}
&E'|_{y=y_{\rm UV},y_{\rm IR}}=0\ ,\\[1ex]
\label{eq:b_ep_i}
&[E']|_{y=y_{\rm I}}=0\ ,\\[1ex]
\label{FGcontinuity}
&[F]|_{y=y_{\rm UV},y_{\rm I},y_{\rm IR}}=[G]|_{y=y_{\rm UV},y_{\rm I},y_{\rm IR}}=0\ .
\end{align}
Eq.\,\eqref{eq:b_phi} and Eq.\,\eqref{eq:b_fg} are also obtained in the two 3-brane system~\cite{Csaki:2000zn}.
The conditions in Eq.\,\eqref{eq:b_ep_f} and Eq.\,\eqref{eq:b_ep_i} were mentioned in~\cite{Pilo:2000et,Kogan:2001qx}
where the three 3-brane system without the GW field is discussed.
Eq.~\eqref{FGcontinuity} comes from the continuity of the metric.

In the bulk spacetime, we obtain the following equations
from the Einstein equation~\eqref{linearized_Einstein_equation} with Eq.~\eqref{deltaRT}
and the field equation~\eqref{linearized_field_equation}:
\begin{align}
\label{eq:1}
&\Box F+e^{-2A}\left(-F''+A''(2F+2G)+A'(8F'+G')+A'^2(-8G-8F)-\frac{1}{2}\Box E' A'\right) \nonumber \\
&\qquad =-\frac{2}{3}\kappa^2e^{-2A}\left(\frac{\partial V}{\partial\phi}\hat\phi-2F V(\phi_0)\right) , \\[1ex]
\label{eq:2}
&2F_{,\mu\nu}-G_{,\mu\nu}+e^{-2A}\left(\frac{1}{2}E''_{,\mu\nu}-A'' E_{,\mu\nu}+4A'^2 E_{,\mu\nu}-2A' E'_{,\mu\nu}\right)
=-\frac{2}{3}\kappa^2e^{-2A} E_{,\mu\nu}V \ ,\\[1ex]
\label{eq:3}
&3 F'_{,\mu}-3A' G_{,\mu}=\kappa^2\partial_\mu\hat\phi\,\phi_0'\ ,\\[1ex]
\label{eq:4}
&e^{2A}\Box G+4F''+A'(-8F'-4G')-\frac{1}{2}\Box E''+A'\Box E'=\kappa^2\left(2\phi_0'\hat\phi'+\frac{2}{3}\left(2GV+\frac{\partial V}{\partial\phi}\hat\phi\right)\right) ,\\[1ex]
\label{eq:5}
&-\hat\phi''+e^{2A}\Box\hat\phi+4A'\hat\phi'+\frac{\partial^2 V}{\partial\phi^2}\hat\phi=-2G\frac{\partial V}{\partial\phi}-(G'+4F')\phi_0'\ .
\end{align}
The first equation \eqref{eq:1} is found by matching the terms that are proportional to
$\eta_{\mu\nu}$ in $\delta R_{\mu\nu}=\kappa^2\delta \widetilde{T}_{\mu\nu}$. 
The second equation \eqref{eq:2} is given by matching the remaining terms. 
The third, fourth and fifth equations \eqref{eq:3}, \eqref{eq:4}, \eqref{eq:5} are obtained, respectively,
via $\delta R_{\mu5}=\kappa^2\delta \widetilde{T}_{\mu5}$, $\delta R_{55}=\kappa^2\delta \widetilde{T}_{55}$
and the field equation for $\phi$.
Eq.~\eqref{eq:2} and Eq.~\eqref{eq:3} are satisfied for
any $_{,\mu\nu}$ and $_{,\mu}$. Then, we find
\begin{align}
\label{eq:G}
&G=2F+e^{-2A}\left(\frac{1}{2}E''-A'' E+4A'^2 E-2A' E'\right)+\frac{2}{3}\kappa^2e^{-2A}EV \ ,\\[1ex]
&\hat\phi=\frac{1}{\kappa^2\,\phi_0'}(3 F'-3A' G)\ ,
\end{align}
which determine $G$ and $\hat \phi$ in terms of the fluctuations $F$ and $E$.
By inserting these expressions for $G$ and $\hat \phi$ into $e^{2A}\times {\rm Eq.}\,\eqref{eq:1}+{\rm Eq.}\,\eqref{eq:4}$,
we obtain
\begin{align}
\label{eq:eq_1}
f''-f'\left(2A'+2\frac{\phi_0''}{\phi_0'}\right)-f\left(4 A''-4A' \frac{\phi_0''}{\phi_0'}\right)= e^{2A}\Box f\ ,
\end{align}
where we redefine $F$ as
\begin{align}
\label{eq:F_redefine}
F(x,y)\equiv f(x,y)+\frac{A'}{2 e^{2A}}E'\ .
\end{align}
Note that Eq.~\eqref{eq:eq_1} does {\it not} include the fluctuation $E$ explicitly and
it has the same form obtained for the two 3-brane system in~\cite{Csaki:2000zn}.
By solving the equation for $f$ with the boundary conditions,
we can compute the KK mass spectrum, radion masses and radion couplings to matter fields
as we will perform in the subsequent sections.
We still do not use the equation for the $\hat\phi$ field in Eq.~\eqref{eq:5},
but it is automatically satisfied by the other equations.
In terms of $f$ and $E$, the boundary condition in Eq.\,\eqref{eq:b_fg} is rewritten as
$f'+({1}/{2})e^{-2A}A'' E'-2 A' f|_{y=y_{\rm UV},y_{\rm I},y_{\rm IR}}=0$.
On the UV and IR branes, $E'|_{y=y_{\rm UV},y_{\rm IR}}=0$ are the boundary conditions, thus we obtain
\begin{align}
\label{boundaryUV}
&f_{1}'-2 A_{1}' f_{1}|_{y=y_{\rm UV}}=0\ , 
\\[1ex]
\label{boundaryIR}
&f_{2}'-2 A_{2}' f_{2}|_{y=y_{\rm IR}}=0\ .
\end{align}
Here, the labels $1,2$ denote functions defined in
the bulk region between the UV and intermediate branes (subregion 1)
and the region between the intermediate and IR branes (subregion 2), respectively.
These conditions are the same as those of the two 3-brane model.
On the other hand, for the intermediate brane, $E'(y_{\rm I})$ is determined
from Eq.\,\eqref{eq:b_ep_i} and Eq.\,\eqref{FGcontinuity}.
By substituting the obtained $E'(y_{\rm I})$ into Eq.\,\eqref{eq:b_fg},
the boundary conditions on the intermediate brane are
\begin{align}
\label{E'continuity}
&{f_{1}'-2A_{1}' f_{1}+\frac{f_1-f_2}{A_2'-A_1'}A_1'' \biggr|_{y=y_{\rm I}}=0\ ,} \\[1ex]
\label{Fcontinuity}
&{f_{2}'-2A_{2}' f_{2}+\frac{f_1-f_2}{A_2'-A_1'}A_2'' \biggr|_{y=y_{\rm I}}=0\ .}
\end{align}
Note that the function $f$ can be discontinuous around the intermediate brane, $y=y_{\rm I}$,
while the fluctuation $F$ is required to be continuous from the continuity of the metric.
A nonzero $E'$ in Eq.~\eqref{eq:F_redefine} on the intermediate brane makes it possible.

In summary, we have obtained a bulk equation for $f$ in each subregion,
\begin{align}
\label{eq:bulk_f_1}
&f_{1}''-f_{1}'\left(2A_{1}'+2\frac{\phi_{0,1}''}{\phi_{0,1}'}\right)-f_{1}\left(4 A_{1}''-4A_{1}' \frac{\phi_{0,1}''}{\phi_{0,1}'}\right)=-m^2 e^{2A_1}f_1 \quad (\text{subregion~$1$}) \ ,  \\[1ex]
\label{eq:bulk_f_2}
&f_{2}''-f_{2}'\left(2A_{2}'+2\frac{\phi_{0,2}''}{\phi_{0,2}'}\right)-f_{2}\left(4 A_{2}''-4A_{2}' \frac{\phi_{0,2}''}{\phi_{0,2}'}\right)=-m^2 e^{2A_2}f_2 \quad (\text{subregion~$2$}) \ ,
\end{align}
where $\Box f_{1,2}=-m^2 f_{1,2}$ have been used.
These equations are solved under the boundary conditions at the orbifold fixed points
\eqref{boundaryUV}, \eqref{boundaryIR} 
and the conditions at the intermediate brane in Eq.\,\eqref{E'continuity} and Eq.\,\eqref{Fcontinuity}.
The above discussion can be generalized to the $N+1$ 3-brane system mentioned in the previous section,
and the bulk equations and the boundary conditions for that case are summarized in appendix~\ref{app:generalN}.

Before closing this section, we comment on one interesting feature of the present problem.
As discussed in ref.~\cite{Csaki:2000zn}, the bulk equation for $f$ can be transformed into the Schr{\"o}dinger-like equation.
By changing the coordinate $dz\,e^{-A(z)} = dy$ with $A (z)=A (y(z))$ and
rescaling the field as $f_s=e^{3/2A_s}\phi_{0,s}' \tilde f_s$ ($s=1,2$), the bulk equations \eqref{eq:bulk_f_1},
\eqref{eq:bulk_f_2} are rewritten as
\begin{align}
\label{eq:schrodinger}
\begin{split}
-\frac{d^2}{dz^2}\tilde f_s+\biggl\{\frac{9}{4} \left(\frac{dA_s}{dz} \right)^2+\frac{5}{2}\frac{d^2A_s}{dz^2}&- \frac{dA_s}{dz}\frac{d^2\phi_{0,s}/dz^2}{d\phi_{0,s}/dz} \\[1ex]
&+2\left(\frac{d^2\phi_{0,s}/dz^2}{d\phi_{0,s}/dz}\right)^2-\frac{d^3\phi_{0,s}/dz^3}{d\phi_{0,s}/dz}\biggr\}\tilde f_s
=m^2 \tilde f_s\ .
\end{split}
\end{align}
This expression has the same form as the Schr{\"o}dinger equation, $\hat O \tilde{f}_s=m^2 \tilde{f}_s$
where $\hat O$ includes derivatives of $\tilde{f}_s$ with respect to $z$.
In appendix~\ref{app:hermiticity}, we derive the following condition that
the hermiticity of the differential operator is satisfied,
\begin{align}
\label{eq:hermiticity}
   \left[\tilde f_{1:n_2} \frac{d}{dz}\tilde f^*_{1:n_1}-\tilde f^*_{1:n_1}\frac{d}{dz}\tilde f_{1:n_2}\right]_{z=z_{\rm UV}}^{z=z_{\rm I}}
     +\left[\tilde f_{2:n_2} \frac{d}{dz}\tilde f^*_{2:n_1}-\tilde f^*_{2:n_1}\frac{d}{dz}\tilde f_{2:n_2}\right]_{z=z_{\rm I}}^{z=z_{\rm IR}}=0\ ,
\end{align}
where $\tilde f_{s:n_1}, \tilde f_{s:n_2}$ denote eigenfunctions for Eq.~\eqref{eq:schrodinger},
$[X(z)]^{z=z_A}_{z=z_B}\equiv X(z_A)-X(z_B)$,
and the integration range is divided into two parts, $i.e.$ the subregions $1,2$.
Let us see if this hermiticity condition is actually satisfied or not.
The boundary conditions in Eq.\,\eqref{eq:b_fg} lead to
\begin{equation}
\begin{split}
\label{eq:b_hermiticity}
&\left.-\tilde f_1 \frac{dA_1}{dz}+2 \frac{d\tilde f_1}{dz}+\frac{2 \tilde f_1 d^2\phi_{0,1}/dz^2}{d\phi_{0,1}/dz}+\frac{e^{-3A_1/2} dE_1/dz((dA_1/dz)^2+d^2A_1/dz^2)}{d\phi_{0,1}/dz}\right|_{z=z_{\rm UV},z_{\rm I}}=0 ,\\[1ex]
&\left.-\tilde f_2 \frac{dA_2}{dz}+2 \frac{d\tilde f_2}{dz}+\frac{2 \tilde f_2 d^2\phi_{0,2}/dz^2}{d\phi_{0,2}/dz}+\frac{e^{-3A_2/2} dE_2/dz((dA_2/dz)^2+d^2A_2/dz^2)}{d\phi_{0,2}/dz}\right|_{z=z_{\rm I},z_{\rm IR}}=0 .
\end{split}
\end{equation}
The hermiticity condition \eqref{eq:hermiticity} is satisfied if $dE_{1,2}/dz=0$ hold on all the branes 
because $d\tilde f_{1,2}/dz$ are, respectively, proportional to $\tilde f_{1,2}$
and the factors of proportionality are the same among different eigenfunctions
as seen via the above boundary conditions. However, $dE_{1,2}/dz$ can be nonzero on the intermediate brane,
and hence the hermiticity condition is not satisfied in general.
It means that the orthogonality between eigenfunctions with different eigenvalues is not guaranteed.
The non-hermiticity of the differential operator has been already discussed in the two 3-brane model~\cite{Csaki:2000zn},
but in this case the non-hermiticity appears for $\hat\phi\neq 0$ on the boundaries,
which is also true for the three 3-brane model.
In Sec.~\ref{sec:radion_mass}, we will encounter the non-hermiticity due to $dE_{1,2}/dz|_{y=y_{\rm I}}\neq 0$
when we obtain two radion solutions and their masses
(the non-hermitian physics is reviewed in $e.g.$ ref.~\cite{Ashida:2020dkc}).

%#######################
\section{KK mass spectra}
\label{sec:KK_towers}
In this section, we compute the mass spectrum of KK modes of the radion-GW scalar system
by solving the bulk equations for $f$ in Eqs.~\eqref{eq:bulk_f_1} and \eqref{eq:bulk_f_2}
with the boundary conditions on the UV and IR branes \eqref{boundaryUV}, \eqref{boundaryIR}
and the conditions on the intermediate brane \eqref{E'continuity}, \eqref{Fcontinuity}.
Here, $\kappa\phi_{\rm UV} \ll 1$ is assumed
and the backreaction effect from the GW scalar field background is neglected.
As we will see in the next section, to calculate radion masses,
it is required to solve the equations at the next-to-leading order of $\kappa\phi_{\rm UV}$.
We also compute the mass spectrum of KK gravitons by
extending the computation in the two 3-brane model~\cite{Csaki:2000zn} to the present system with three 3-branes.

\subsection{The radion-scalar system}
\label{radionscalar}

Our strategy is to solve the bulk equation for each of the subregions $1,2$ with the UV or IR boundary condition
and then match the two solutions for the subregions $1,2$ on the intermediate brane
by taking account of the boundary conditions.
In the subregion $1$ between the UV and intermediate branes, neglecting the GW scalar field background
and using the explicit solution of $A(y)$ in Eq.~\eqref{backgroundsolutions},
the bulk equation \eqref{eq:bulk_f_1} for $f_1$ is reduced to
\begin{align}
f_1''- 2(k_1+u_1)f_1' +4k_1 u_1f_1=-m^2 e^{2k_1 y}f_1 \ ,
\end{align}
and the boundary condition on the UV brane \eqref{boundaryUV} is
\begin{align}
   f_1'(0)-2k_1 f_1(0)=0\ .
   \label{eq:b_f1_UV}
\end{align}
Then, the solution is found as
\begin{equation}
\begin{split}
\label{f1sol}
f_1 (y)=c_1\, e^{y (k_1+u_1)} \Biggl[&J_{1-\frac{u_1}{k_1}}\left(\mathcal{Z}_{k_1}(y)\right)
\Bigl\{m
   J_{\frac{u_1}{k_1}}\left(\mathcal{Z}_{k_1}(0)\right)-m
   J_{\frac{u_1}{k_1}-2}\left(\mathcal{Z}_{k_1}(0)\right) \\
&+2 (k_1-u_1)
   J_{\frac{u_1}{k_1}-1}\left(\mathcal{Z}_{k_1}(0)\right) \Bigr\} -2 m J_{2-\frac{u_1}{k_1}}\left(\mathcal{Z}_{k_1}(0)\right)
   J_{\frac{u_1}{k_1}-1}\left(\mathcal{Z}_{k_1}(y)\right)\Biggr]\ ,
\end{split}
\end{equation}
where $c_1$ is a constant, $J_n(\cdot)$ represents the Bessel function of the first kind of order $n$,
and we have defined 
\begin{align}
    \mathcal{Z}_{k_1}(y)\equiv\frac{e^{k_1y}m}{k_1} \ ,
    \qquad \mathcal{Z}_{k_2}(y)\equiv\frac{e^{k_2(y-y_{\rm I})+k_1y_{\rm I}}m}{k_2}\ .
\end{align}
On the other hand, in the subregion $2$ between the intermediate and IR branes,
the bulk equation for $f_2$ in Eq.~\eqref{eq:bulk_f_2} and the boundary condition on the IR brane
\eqref{boundaryIR} are, respectively, given by
\begin{align}
&f_2''- 2(k_2+u_2)f_2'+4k_2 u_2f_2=-m^2 e^{2k_2 y}f_2 \ ,\\[1ex]
&f_2'(y_{\rm IR})-2k_2 f_2(y_{\rm IR})=0\ .
\label{eq:b_f2_IR}
\end{align}
The solution is obtained as
\begin{align}
\label{f2sol}
f_2 (y)=&\, c_2\,e^{y (k_2+u_2)-k_2 y_{\rm I}} 
\biggl[m\, e^{k_1 y_{\rm I}+k_2 y_{\rm IR}}
   J_{1-\frac{u_2}{k_2}}\left(\mathcal{Z}_{k_2}(y)\right) J_{\frac{u_2}{k_2}}\left(\mathcal{Z}_{k_2}(y_{\rm IR})\right) \nonumber \\[1ex]
&+J_{1-\frac{u_2}{k_2}}\left(\mathcal{Z}_{k_2}(y)\right) \left\{2 (k_2-u_2) e^{k_2
   y_{\rm I}} J_{\frac{u_2}{k_2}-1} \left(\mathcal{Z}_{k_2}(y_{\rm IR})\right)-m\, e^{k_1 y_{\rm I}+k_2 y_{\rm IR}}
   J_{\frac{u_2}{k_2}-2}\left(\mathcal{Z}_{k_2}(y_{\rm IR})\right)\right\}\nonumber\\[1ex]
   &-2 m\, e^{k_1 y_{\rm I}+k_2 y_{\rm IR}}
   J_{\frac{u_2}{k_2}-1}\left(\mathcal{Z}_{k_2}(y)\right) J_{2-\frac{u_2}{k_2}}\left(\mathcal{Z}_{k_2}(y_{\rm IR})\right)\biggr]
\end{align}
where $c_2$ is a constant.

\begin{figure*}
\begin{minipage}[t]{\hsize}
\includegraphics[width=7cm]{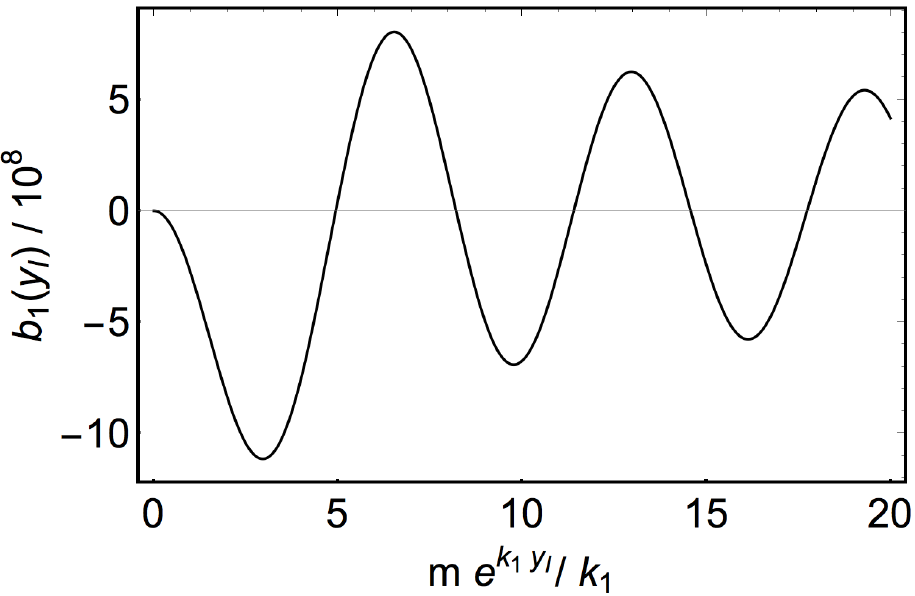}
\hspace{1cm}
\includegraphics[width=7cm]{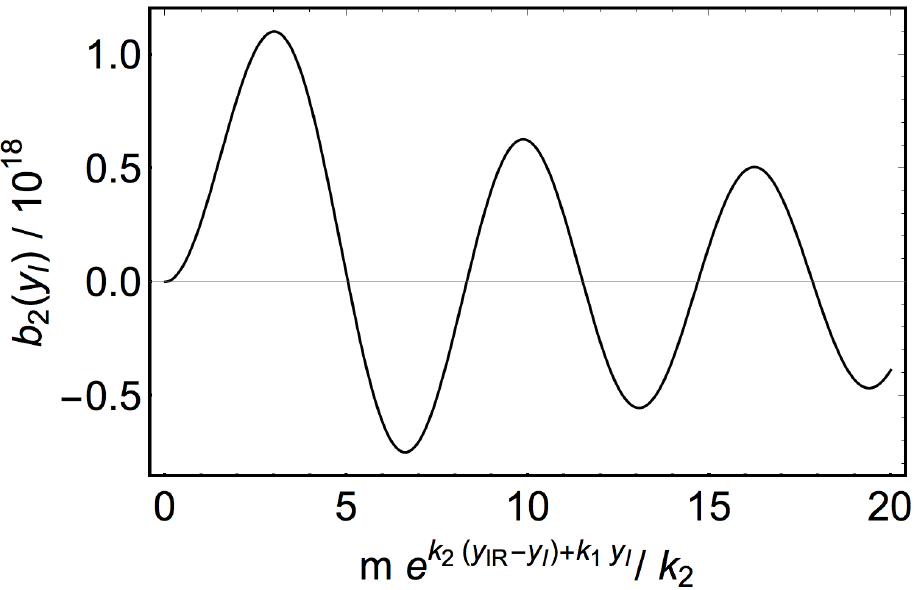}
  \end{minipage}
  \caption{The normalized $b_1 (y_{\rm I})$ in Eq.~\eqref{eq:KK_b_2} and $b_2 (y_{\rm I})$ in Eq.~\eqref{eq:KK_b_3}
  as functions of the normalized mass parameters,
  $m\,e^{k_1y_{\rm I}}/k_1$ and $m\,e^{k_2(y_{\rm IR}-y_{\rm I})+k_1y_{\rm I}}/k_2$, respectively.
  Here, we take 
    $k_1y_{\rm I}=10,~k_2y_{\rm I}=20,~k_2y_{\rm IR}=40,~u_1/k_1=0.15,~u_2/k_2=0.075$
  as reference values.
}
\label{fig:b1b2} 
\end{figure*}

Having obtained the solutions \eqref{f1sol}, \eqref{f2sol} for the two subregions,
let us connect them by using the boundary conditions at the intermediate brane \eqref{E'continuity}, \eqref{Fcontinuity}
which are rewritten as
\begin{align}
\label{eq:KK_b_2}
&b_1(y_{\rm I})\equiv f_1'(y_{\rm I})-2
k_1f_1(y_{\rm I})
=0\ ,\\[1ex]
\label{eq:KK_b_3}
&b_2(y_{\rm I})\equiv f_2'(y_{\rm I})-2
k_2
f_2(y_{\rm I})
=0\ .
\end{align}
Here, we have ignored the backreaction effect from the GW scalar field background. 
For massive modes with $m>0$, Eq.~\eqref{eq:KK_b_3} can be satisfied with $f_2(y)=0$ ($c_2 = 0$ in Eq.~\eqref{f2sol})
which indicates that $f_2$ is zero for the whole subregion $2$.
The condition \eqref{eq:KK_b_2} then determines $m$.
The left panel of Fig.~\ref{fig:b1b2} shows $b_1(y_{\rm I})$ as a function of $m$.
We can see from the figure that the lowest KK excitation has $m\sim k_1 e^{-k_1 y_{\rm I}}$,
and the masses for higher KK excitations are also found.
The other massive modes are obtained by taking $f_1(y)=0$ ($c_1 = 0$ in Eq.~\eqref{f1sol})
for the whole subregion $1$ to satisfy Eq.~\eqref{eq:KK_b_2}.
In this case, the condition \eqref{eq:KK_b_3} leads to masses around
$m\sim k_2 e^{-k_2 y_{\rm IR}+y_{\rm I}(k_2-k_1)}$,
as seen in the right panel of Fig.~\ref{fig:b1b2}.
Therefore, we have obtained a set of eigenfunctions $f$,
which are nonzero in only one bulk subregion $p$ $(= 1 \, \text{or} \, 2)$,
with mass eigenvalues $m\sim k_p e^{-A(y_p)}$.
Note that these eigenfunctions for the KK modes with different mass eigenvalues are orthogonal to each other
because the hermiticity condition in Eq.\,\eqref{eq:hermiticity} is satisfied with the boundary conditions
in Eqs.\,\eqref{eq:b_f1_UV}, \eqref{eq:b_f2_IR}, \eqref{eq:KK_b_2}, \eqref{eq:KK_b_3}.

Zero mode solutions to the bulk equations with $m=0$ are given by $f_p\propto e^{2k_p y}$
in a subregion $p$ $(= 1 \, \text{or} \, 2)$ and $f_{p'}=0$ in the other subregion $p'$ $(= 2 \, \text{or} \, 1)$.
The two solutions automatically satisfy the boundary conditions at the intermediate brane
\eqref{eq:KK_b_2}, \eqref{eq:KK_b_3}
as well as the conditions at the UV and IR branes. 
These zero modes are understood as two radion degrees of freedom
which correspond to the fluctuations of two branes ($e.g.$ the IR and intermediate branes)
relative to the third brane ($e.g.$ the UV brane).
As we will see in the next section, they become massive once the backreaction effect
from the GW scalar field background is included,
and the mass eigenstates are given by linear combinations of the two solutions where the backreaction effect is ignored.

\subsection{KK gravitons}

We can introduce spin-$2$ fluctuations, $i.e.$ gravitons, by changing the metric of Eq.~\eqref{metric},
$\eta_{\mu\nu}\to \eta_{\mu\nu}+e^{2A(y)}h_{\mu\nu}$,
where $h_{\mu\nu}=h_{\nu\mu}$ is a symmetric tensor satisfying the transverse
($\eta^{\mu\nu}\partial_\mu h_{\nu\rho}=0$) and traceless ($\eta^{\mu\nu}h_{\mu\nu}=0$) conditions.
In the linearized Einstein equation for the fluctuations,
the tensor modes do not mix with the spin-$0$ modes of $F,G,E$ and $\hat{\phi}$.
By ignoring the backreaction effect from the GW scalar field background,
the bulk equations for the tensor modes are obtained via $\delta R_{\mu\nu}=\kappa^2\delta \widetilde T_{\mu\nu}$
in the subregions $1$ and $2$, respectively,
\begin{align}
\label{hbulk1}
&h_{1\mu\nu}^{''}-4k_1^2h_{1\mu\nu}+m^2 e^{2k_1 y}h_{1\mu\nu}=0\ ,\\[1ex]
\label{hbulk2}
&h_{2\mu\nu}^{''}-4k_2^2h_{2\mu\nu}+m^2 e^{2k_2 y+2y_{\rm I}(k_1-k_2)}h_{2\mu\nu}=0\ .
\end{align}
Here, the subscripts $1,2$ indicate the functions in the subregions $1,2$
and $\Box h_{p\mu\nu}=-m^2h_{p\mu\nu}$ $(p=1,2)$ have been used.
The prime $'$ denotes a derivative with respect to $y$.
The other equations, $\delta R_{\mu 5}=\kappa^2\delta \widetilde T_{\mu5}$ and $\delta R_{55}=\kappa^2\delta \widetilde T_{55}$, are automatically satisfied for the transverse and traceless tensor modes~\cite{Charmousis:1999rg}. 
The boundary conditions on the UV, intermediate and IR branes are obtained
through $\delta R_{\mu\nu}=\kappa^2\delta \widetilde T_{\mu\nu}$,
matching the singularities on the branes,
\begin{align}
\label{eq:b_h_UV}
&h^{'}_{1\mu\nu}+2k_1h_{1\mu\nu}|_{y=y_{\rm UV}}=0\ ,\\[1ex]
\label{eq:b_h_I}
&h^{'}_{1\mu\nu}-h^{'}_{2\mu\nu}+2(k_1-k_2)h_{1\mu\nu}|_{y=y_{\rm I}}=0\ ,\\[1ex]
\label{eq:b_h_IR}
&h^{'}_{2\mu\nu}+2k_2h_{2\mu\nu}|_{y=y_{\rm IR}}=0\ ,
\end{align}
which are equivalent to the Israel junction conditions.
The first and third conditions are the same as those of the two 3-brane model~\cite{Charmousis:1999rg}.
We also require the continuity of the tensor around the intermediate brane,
\begin{align}
\label{hcontinuity}
\left.h_{1\mu\nu}-h_{2\mu\nu}\right|_{y=y_{\rm I}}=0\ .
\end{align}
In the following, we take an ansatz $h_{1,2\,\mu\nu}(x,y)=h_{1,2}(y)\chi_{\mu\nu}(x)$,
where $\chi_{\mu\nu}(x)$ satisfies the transverse and traceless conditions,
and solve the equations to determine $h_{1,2}(y)$.

We first solve the equations for $m=0$.
The bulk equation \eqref{hbulk1} for the subregion 1 and the boundary condition at the UV brane \eqref{eq:b_h_UV} lead to
\begin{align}
h_1 (y)=c_{0,h_1}e^{-2k_1y}\ ,
\end{align}
where $c_{0,h_1}$ is a constant.
On the other hand, the bulk equation \eqref{hbulk2} for the subregion 2
and the boundary condition at the IR brane \eqref{eq:b_h_IR} give
\begin{align}
h_2 (y) =c_{0,h_2}e^{-2k_2y-2y_{\rm I}(k_1-k_2)}\ ,
\end{align}
where $c_{0,h_2}$ is a constant.
The continuity condition around the intermediate brane \eqref{hcontinuity},
$i.e.$ $[h]|_{y=y_{\rm I}}=0$, requires $c_{0,h_1}=c_{0,h_2}$.
In this case, the boundary condition at the intermediate brane \eqref{eq:b_h_I} is automatically satisfied. 
The field value of the zero mode is exponentially suppressed further away from the UV brane. 

\begin{figure}
\includegraphics[width=7cm]{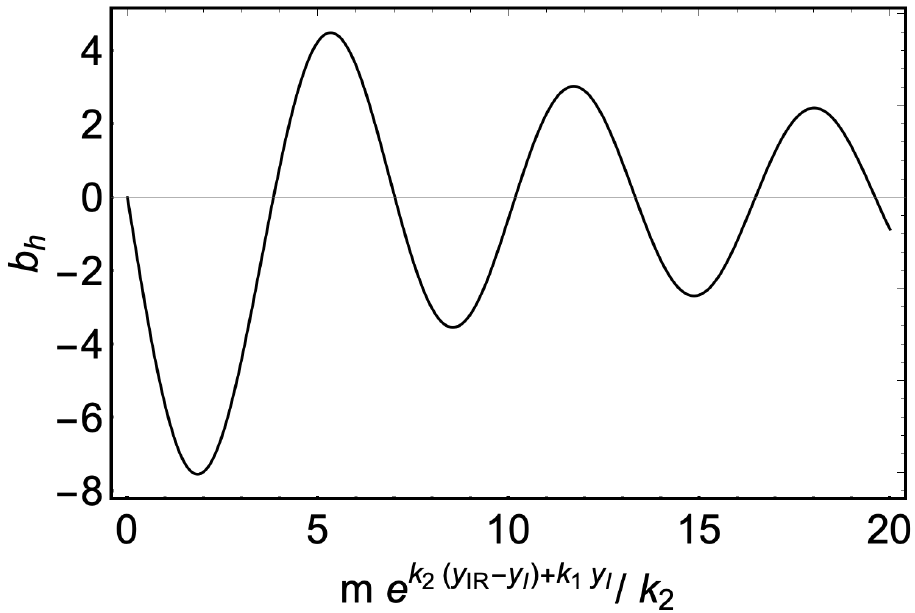}
  \caption{The function $b_h$ defined in Eq.~\eqref{bh} as a function of the normalized mass parameter
  $m\,e^{k_2(y_{\rm IR}-y_{\rm I})+k_1y_{\rm I}}/k_2$.
  The KK graviton masses are determined via $b_h=0$. 
  Here, we take 
  $k_1y_{\rm I}=10,~k_2y_{\rm I}=20,~k_2y_{\rm IR}=40$
  as reference values.
  The first excitation mode is obtained around $m\sim 4k_2e^{-k_2(y_{\rm IR}-y_{\rm I})-k_1y_{\rm I}}$.
}
\label{fig:KKgraviton} 
\end{figure}

Next, let us discuss massive tensor modes.
By solving the bulk equation \eqref{hbulk1} and the boundary condition at the UV brane \eqref{eq:b_h_UV},
we obtain
\begin{align}
h_1 (y)=\,\,&2 c_{m,h_1} \biggl[ \, J_2\left(\mathcal{Z}_{k_1} (y)\right)-\frac{J_1\left(\mathcal{Z}_{k_1} (0)\right)
   Y_2\left(\mathcal{Z}_{k_1} (y)\right)}{Y_1\left(\mathcal{Z}_{k_1} (0)\right)} \, \biggr]\ ,
\end{align}
for the subregion 1.
Here, $Y_n(\cdot)$ denotes the Bessel function of the second kind of order $n$ and $c_{m,h_1}$ is a constant.
On the other hand, we solve the bulk equation \eqref{hbulk2} for the subregion 2
with the IR boundary condition \eqref{eq:b_h_IR} and the connection $[h]|_{y=y_{\rm I}}=0$ and get
\begin{align}
h_2 (y)=\,\,&2 c_{m,h_2} \biggl[ \, J_2\left(\mathcal{Z}_{k_2} (y)\right)-\frac{J_1\left(\mathcal{Z}_{k_2} (y_{\rm IR})\right)Y_2\left(\mathcal{Z}_{k_2} (y)\right)}{Y_1\left(\mathcal{Z}_{k_2} (y_{\rm IR})\right)} \, \biggr] \ ,
\end{align}
where $c_{m,h_2}$ and $c_{m,h_1}$ are related as
\begin{align}
\begin{split}
c_{m,h_2}=\,\, &c_{m,h_1}\Bigl[\Bigl\{Y_1\left(\mathcal{Z}_{k_1} (0)\right)
   J_2\left(\mathcal{Z}_{k_1} (y_{\rm I})\right)-J_1\left(\mathcal{Z}_{k_1} (0)\right)
   Y_2\left(
\mathcal{Z}_{k_1} (y_{\rm I})\right)\Bigr\}
   Y_1\left(\mathcal{Z}_{k_2} (y_{\rm IR})\right)\Bigr]\\[1ex]
& \Bigl/ \Bigr[Y_1\left(\mathcal{Z}_{k_1} (0)\right)
   \Bigr\{J_2\left(\mathcal{Z}_{k_2} (y_{\rm I})\right) Y_1\left(\mathcal{Z}_{k_2} (y_{\rm IR})\right)-Y_2\left(\mathcal{Z}_{k_2} (y_{\rm I})\right)
   J_1\left(\mathcal{Z}_{k_2} (y_{\rm IR})\right)\Bigr\}\Bigr]\ .
\end{split}
\end{align}
The last equation to be satisfied is the junction condition on the intermediate brane in Eq.\,\eqref{eq:b_h_I},
\begin{align}
\label{bh}
0=\,\,&J_1\left(\mathcal{Z}_{k_2} (y_{\rm IR})\right) \Bigl[Y_1\left(\mathcal{Z}_{k_1} (0)\right)
   J_2\left(\mathcal{Z}_{k_1} (y_{\rm I})\right)
   Y_1\left(\mathcal{Z}_{k_2} (y_{\rm I})\right) \nonumber \\[1ex]
   &+Y_2\left(\mathcal{Z}_{k_2} (y_{\rm I})\right) \Bigl\{J_1\left(\mathcal{Z}_{k_1} (0)\right)
   Y_1\left(\mathcal{Z}_{k_1} (y_{\rm I})\right)-Y_1\left(\mathcal{Z}_{k_1} (0)\right)
   J_1\left(\mathcal{Z}_{k_1} (y_{\rm I})\right)\Bigr\} \nonumber \\[1ex]
   &-J_1\left(\mathcal{Z}_{k_1} (y_{\rm I})\right)
   Y_2\left(\mathcal{Z}_{k_1} (y_{\rm I})\right)
   Y_1\left(\mathcal{Z}_{k_2} (y_{\rm I})\right)\Bigr] \nonumber \\[1ex]
   &+Y_1\left(\mathcal{Z}_{k_2} (y_{\rm IR})\right)
   \Bigl[Y_1\left(\mathcal{Z}_{k_1} (0)\right) J_1\left(\mathcal{Z}_{k_1} (y_{\rm I})\right) J_2\left(\mathcal{Z}_{k_2} (y_{\rm I})\right) \nonumber \\[1ex]
   &-J_1\left(\mathcal{Z}_{k_1} (0)\right)
   Y_1\left(\mathcal{Z}_{k_1} (y_{\rm I})\right)
   J_2\left(\mathcal{Z}_{k_2} (y_{\rm I})\right) \nonumber \\[1ex]
   &+J_1\left(\mathcal{Z}_{k_2} (y_{\rm I})\right) \Bigl\{J_1\left(\mathcal{Z}_{k_1} (0)\right)
   Y_2\left(\mathcal{Z}_{k_1} (y_{\rm I})\right)-Y_1\left(\mathcal{Z}_{k_1} (0)\right)
   J_2\left(\mathcal{Z}_{k_1} (y_{\rm I})\right)\Bigr\}\Bigr]\equiv \frac{b_{h}(m)}{2e^{k_1y_{\rm I}}m^2}\ .
\end{align}
The masses of KK excitation states are determined by the values of $m$ where the function $b_h(m)$ crosses zero. 
This function has the following form,
\begin{align}
\begin{split}
b_{h}(m)=\,\, &J_1\left(\mathcal{Z}_{k_2} (y_{\rm IR})\right) a_1\left(\mathcal{Z}_{k_1} (0),\mathcal{Z}_{k_1} (y_{\rm I}),\mathcal{Z}_{k_2} (y_{\rm I})\right) \\[1ex]
&+Y_1\left(\mathcal{Z}_{k_2} (y_{\rm IR})\right) a_2\left(\mathcal{Z}_{k_1} (0),\mathcal{Z}_{k_1} (y_{\rm I}),\mathcal{Z}_{k_2} (y_{\rm I})\right) \ ,
\end{split}   
\end{align}
where $a_{1,2}(\,\cdot\, ,\,\cdot \, ,\,\cdot\,)$ are oscillating functions of $m$.
In this expression,
$J_1\left(\mathcal{Z}_{k_2} (y_{\rm IR})\right)$ and $Y_1\left(\mathcal{Z}_{k_2} (y_{\rm IR})\right)$
are the most rapidly oscillating, which indicates that the mass of the first KK excitation state is determined around $\mathcal{Z}_{k_2} (y_{\rm IR})\sim 1$, $i.e.$ $m\sim k_2 e^{-k_2(y_{\rm IR}-y_{\rm I})-k_{\rm I}y_{\rm I}}$.
In fact, Fig.~\ref{fig:KKgraviton} shows $b_h$ as a function of the normalized mass parameter
$m\,e^{k_2(y_{\rm IR}-y_{\rm I})+k_1y_{\rm I}}/k_2$ and that it is actually the case.
Thus, 
the first few excited states have masses comparable to the typical mass scale of the IR brane.
The similar result was discussed in~\cite{Kogan:2000xc}.

%#######################
\section{Masses of radions}
\label{sec:radion_mass}

In section~\ref{radionscalar}, it was found that two independent zero mode solutions for spin-0 fluctuations,
which correspond to two radions in our three 3-brane setup.
We here take account of the backreaction effect from the GW scalar field background
and find nonzero radion masses.
In the same way as what we did to get the masses of KK modes,
we first solve the bulk equations in the two subregions with the boundary conditions on the UV and IR branes.
Then, the two solutions in different regions are connected by the boundary conditions on the intermediate brane,
which determines the radion masses.
We will see that one radion mass is close to the typical mass scale of the IR brane
and another is at the order of the typical mass scale of the intermediate brane.
Radion couplings to brane-localized matter fields will be discussed in the next section.

Let us expand the fluctuation $f_1$ defined in Eq.~\eqref{eq:F_redefine} for the subregion 1 between the UV and
intermediate branes in terms of $l^2\equiv\kappa^2 \phi_{\rm UV}^2 / 2$ and rewrite it as
\begin{align}
\label{f1expression}
f_1 (y)\equiv c_{f1} e^{2k_1 y}+l^2 e^{2k_1 y}d_1(y)\ ,
\end{align}
where the first term in the right hand side represents the solution at the $0$-th order of $l^2$
with an undetermined constant $c_{f1}$ and the second term gives a perturbation. 
By substituting this expression into the bulk equation in Eq.\,\eqref{eq:bulk_f_1},
the $0$-th order equation is automatically satisfied, and we obtain the following equation at the first order of $l^2$,
\begin{align}
\label{d1bulk}
3 e^{2u_1 y}d_1''(y)+6 e^{2u_1 y}(k_1+u_1)d_1'(y)+4c_{f1}(k_1-u_1)u_1+3c_{f1}e^{2(k_1+u_1)y}m^2=0\ .
\end{align}
Here, we have redefined $m^2 \rightarrow l^2 m^2$. 
Note that this equation does not include terms proportional to $d_1$
and only includes terms with derivatives with respect to $y$, $i.e.$ $d_1'$ and $d_1''$.
The boundary condition on the UV brane in Eq.\,\eqref{boundaryUV} is given by
\begin{align}
\label{d1UV}
2c_{f1}u_1+3
d_1'(0) =0
\ .
\end{align}
The solution to the equations \eqref{d1bulk}, \eqref{d1UV} is obtained as
\begin{align}
\begin{split}
\label{eq:f_1p}
d_1' (y)=-\frac{c_{f1} e^{-2 y (k_1+u_1)}}{6 k_1 (2 k_1+u_1)} \, \Bigl[&\, 8 k_1^2 u_1 e^{2 k_1 y}+3 k_1 m^2 \left(e^{2 y
   (2 k_1+u_1)}-1\right) \\
   &-4 u_1^3 \left(e^{2 k_1 y}-1\right)-4 k_1 u_1^2 \left(e^{2 k_1
   y}-2\right)  \Bigr]\ . 
\end{split}   
\end{align}
On the other hand, for the subregion 2 between the intermediate and IR branes, we rewrite $f_2$ as
\begin{align}
\label{f2expression}
f_2 (y)\equiv c_{f2} e^{2(k_2y+y_{\rm I}(k_1-k_2))}+l^2 e^{2(k_2y+y_{\rm I}(k_1-k_2))} d_2(y)\ .
\end{align}
The first term represents the solution at the $0$-th order of $l^2$
with a constant $c_{f2}$ and the second term gives a perturbation. 
By substituting this expression into Eq.\,\eqref{eq:bulk_f_2},
the $0$-th order equation is automatically satisfied, and we find
\begin{align}
\begin{split}
\label{d2bulk}
&3 c_{f2} m^2 e^{2 (k_2 (y-y_{\rm I})+y_{\rm I} (k_1+u_1)+u_2 y)}-4 c_{f2} u_2(u_2-k_2) e^{2 u_2 y_{\rm I}}\\[1ex]
&\quad \qquad+3 d_2''(y) e^{2 u_2 y+2 u_1 y_{\rm I}}+6 (k_2+u_2)
   d_2'(y) e^{2 u_2 y+2 u_1 y_{\rm I}}=0\ ,
\end{split}   
\end{align}
at the first order of $l^2$.
This is also the first-order differential equation for $d_2'$ as in the case of $d_1'$.
The boundary condition on the IR brane in Eq.\,\eqref{boundaryIR} is
\begin{align}
\label{d2IR}
2 c_{f2} u_2 
+3 d_2'(y_{\rm IR}) e^{2 y_{\rm I}
   u_1
   +2 u_2 (y_{\rm IR}-y_{\rm I})} =0
\ .
\end{align}
We solve the equations \eqref{d2bulk}, \eqref{d2IR} as
\begin{align}
\label{eq:f_2p}
\begin{split}
d_2' (y)= \,\, 
   &-\frac{c_{f2}\, e^{-2 (k_2 (y+y_{\rm I})+u_2 (y+y_{\rm IR})+u_1 y_{\rm I})}}{6 k_2 (2 k_2+u_2)} \,
   \Bigl[ \, 3 k_2
   m^2 e^{4 k_2 y+2 y_{\rm I} (k_1+u_1)+2 u_2 (y+y_{\rm IR})} \\[1ex] 
   &\qquad -4 u_2 \left(-2
   k_2^2+k_2 u_2+u_2^2\right) e^{2 k_2 (y+y_{\rm I})+2 u_2 (y_{\rm I}+y_{\rm IR})} \\[1ex]
   &\qquad -3 k_2 m^2 e^{4 y_{\rm IR}
   (k_2+u_2)+2 y_{\rm I} (k_1 +u_1)}+4 u_2^2 (2 k_2+u_2) e^{2
   (k_2+u_2) (y_{\rm I}+y_{\rm IR})}\Bigr]\ .
\end{split}   
\end{align}
Next, let us use the boundary conditions on the intermediate brane in Eq.\,\eqref{E'continuity} and  Eq.\,\eqref{Fcontinuity} that are rewritten in terms of $d_1'(y_{\rm I})$ and $d_2'(y_{\rm I})$,
\begin{align}
\label{eq:B_1_yI}
&
-\frac{2u_1^2(c_{f1}-c_{f2})
}{k_1-k_2}+
2c_{f1}u_1+3e^{2u_1y_{\rm I}}d_1'(y_{\rm I})
=0\ ,\\[1ex]
&-\frac{2u_2^2(c_{f1}-c_{f2})
}{k_1-k_2} 
+
2 c_{f2} u_2 
+3 e^{
2 u_1 y_{\rm I}
   }d_2'(y_{\rm I}) 
   =0\ .
   \label{eq:B_2_yI}
\end{align}
By substituting $d_1'(y_{\rm I})$ in Eq.\,\eqref{eq:f_1p} into the condition \eqref{eq:B_1_yI}, we get
\begin{align}
\begin{split}
\label{eq:cR_cL}
c_{f1}=\,\,&4 c_{f2} k_1 u_1^2 (2 k_1+u_1) e^{2 k_1 y_{\rm I}}\\[1ex]
&\!\times\Bigl[\, k_2 \Bigl\{-3 k_1 m^2
   \left(e^{2 y_{\rm I} (2 k_1+u_1)}-1\right)+4 u_1^3 \left(e^{2 k_1 y_{\rm I}}-1\right)+8 k_1 u_1^2
   \left(e^{2 k_1 y_{\rm I}}-1\right)\Bigr\} \\[1ex]
   &\qquad +k_1 \Bigl\{3 k_1 m^2 \left(e^{2 y_{\rm I} (2
   k_1+u_1)}-1\right)
   +8 k_1 u_1^2+4 u_1^3 \, \Bigr\} \, \Bigr]^{-1}\ .
\end{split}   
\end{align}
Note that $c_{f1}=c_{f2}$ is satisfied for $k_1=k_2$.
By using 
Eq.\,\eqref{eq:f_2p} and Eq.\,\eqref{eq:cR_cL}, 
the condition \eqref{eq:B_2_yI} leads to
\begin{align}
 \label{eq:exact_masssq}
    \alpha\,m^4+\beta\,m^2+\gamma=0\ ,
\end{align}
where
\begin{align}
\begin{split}
\label{alphabetagamma}
&\alpha=9 k_1 k_2 (k_2-k_1) \left(e^{2 y_{\rm I} (2
   k_1+u_1)}-1\right) \left(e^{2 y_{\rm I} (3 k_1+2
   k_2+u_1+u_2)}-e^{2y_{\rm I}(3k_1 +u_1 ) + 2 y_{\rm IR} (2k_2 + u_2)}\right) ,\\[2ex]
&\beta= 12  \Bigl[k_2
   u_1^2 (2 k_1+u_1) \left(k_2 \left(e^{2 k_1
   y_{\rm I}}-1\right)+k_1\right) \left(e^{2y_{\rm I}(3k_1 +u_1 ) + 2 y_{\rm IR} (2k_2 + u_2)}-e^{2 y_{\rm I} (3 k_1+2
   k_2+u_1+u_2)}\right) \\[1ex]
   &\qquad \quad \left.+k_1 u_2^2 (k_1-k_2) (2
   k_2+u_2) \left(e^{2 y_{\rm I} (2 k_1+u_1)}-1\right) \left(e^{2
   k_2 y_{\rm I}}-e^{2 k_2 y_{\rm IR}}\right) e^{2 y_{\rm I} (2
   k_1+k_2+u_2)}\right.\\[1ex]
   &\qquad \quad +k_1 k_2 u_2^2 (2
   k_2+u_2) \left(e^{2 y_{\rm I} (2 k_1+u_1)}-1\right) e^{2
   y_{\rm I} (2 (k_1+k_2)+u_2)}\Bigr]\,  , \\[2ex]
&\gamma=16 u_1^2 u_2^2 (2
   k_1+u_1) (2 k_2+u_2) e^{2 y_{\rm I} (2
   k_1+k_2+u_2)} \Bigl[k_1 e^{2 k_2 y_{\rm I}}-e^{2
   k_2 y_{\rm IR}} \left(k_2 \left(e^{2 k_1
   y_{\rm I}}-1\right)+k_1\right)\Bigr] \,  .
 \end{split}  
\end{align}
This is a quadratic equation for $m^2$ to have two solutions corresponding to the two radion masses.

Before solving the equation~\eqref{eq:exact_masssq} with the coefficients~\eqref{alphabetagamma} numerically,
we first derive approximate analytical expressions for the two radion masses in a small backreaction limit.
They are obtained by solving Eq.\,\eqref{eq:B_2_yI} with the use of Eq.\,\eqref{eq:f_2p} and Eq.\,\eqref{eq:cR_cL}.
We here assume $e^{k_2y_{\rm IR}}\gg e^{k_1y_{\rm I}}\gg 1$ and $m^2\ll k_1^2 e^{-2k_1 y_{\rm I}}$
and obtain the following approximated expressions for 
$d'_2(y_{\rm I})$ in Eq.\,\eqref{eq:f_2p} and $c_{f2}$ from Eq.\,\eqref{eq:cR_cL}, 
\begin{align}
\label{eq:d2_yi}
&d_2'(y_{\rm I})\simeq-\frac{c_{f2}e^{-2(2k_2+u_2+u_1)y_{\rm I}}}{6k_2(2k_2+u_2)} \Bigl[
3e^{2(k_1+u_1)y_{\rm I}+2(2k_2+u_2)y_{\rm IR}}k_2m^2 \nonumber \\
&\qquad \qquad \qquad \qquad \qquad \qquad \quad -4e^{2u_2y_{\rm I}+2k_2(y_{\rm I}+y_{\rm IR})}u_2^2(2k_2+u_2)\Bigr]\ ,\\[2ex]
\label{eq:cf2}
&c_{f2}\simeq c_{f1}\frac{3e^{2(k_1+u_1)y_{\rm I}} m^2k_1(k_1-k_2)
+4k_2u_1^2(2
k_1+
u_1)}{4
k_1u_1^2(2k_1+u_1)}\ .
\end{align}
We assume that the first and second radion masses are around the typical energy scales of the intermediate and IR branes, respectively.
This assumption will be justified later when the analytical results are compared to the numerical computations.
Let us first find the analytic expression for the radion mass around the IR brane energy scale.
We consider the parameter space $k_1\sim k_2\gtrsim u_2\sim u_1$ for simplicity.
For $c_{f2}$ in Eq.\,\eqref{eq:cf2}, by using $m^2\ll k_1^2 e^{-2k_1 y_{\rm I}}$ and neglecting the term with $m^2$ in the numerator,
the even simpler form is obtained,
\begin{align}
c_{f2}^{\rm IR} \simeq c_{f1}^{\rm IR}\frac{k_2}{k_1}
\ ,
\end{align}
where the dependence on $m^2$ vanishes compared to Eq.\,\eqref{eq:cf2}
and the label ${\rm IR}$ is used to denote the lighter radion.
By substituting $c_{f2}^{\rm IR}$ and Eq.\,\eqref{eq:d2_yi} into Eq.\,\eqref{eq:B_2_yI},
we find a linear equation for $m^2$ leading to
\begin{align}
\label{eq:m_radion_2}
m_{r_{\rm IR}}^2 \simeq \frac{4}{3}\, l^2 u_2^2 \left(2+\frac{u_2}{k_2} \right) e^{-2((k_2+u_2)y_{\rm IR}+(k_1-k_2+u_1-u_2)y_{\rm I}
)}\ .
\end{align}
Here, we have recovered $l^2$ in $m^2$.
This formula is quite similar to the one obtained in the two 3-brane model~\cite{Csaki:2000zn} while the exponential factor is different.
The radion mass is suppressed compared to the typical mass scale of the IR brane
$\sim k_2\, e^{-k_2(y_{\rm IR}-y_{\rm I})-k_1 y_{\rm I}}$
by a factor of about $l e^{-u_2y_{\rm IR}-y_{\rm I}(u_1-u_2)} ({u_2}/{k_2})$. 
In appendix~\ref{app:naive_ansatz}, we also compute the radion mass by using the naive ansatz for the radion field,
where the radion mass-squared is proportional to $(u_2/k_2)^{3/2}$
while the above $m^2_{r_{\rm IR}}/k_2^2$ is proportional to $(u_2/k_2)^{2}\left(2+{u_2}/{k_2} \right)$.
In more detail, the factor in the mass-squared is different by about $ e^{-2u_2y_{\rm IR}-2y_{\rm I}(u_1-u_2)} ({u_2}/{k_2})^{1/2}$.

Next, we shall discuss the radion mass around the intermediate brane.
In this mass range, the first term in Eq.\,\eqref{eq:d2_yi} is dominant, and $d_2'(y_{\rm I})$ is reduced to
\begin{align}
d_2^{{\rm I}'}(y_{\rm I}) \simeq \frac{c^{\rm I}_{f2} }{2(2k_2+u_2)}m^2e^{2((2k_2+u_2)y_{\rm IR}
-(2k_2-k_1+u_2)y_{\rm I})}\ ,
\end{align}
where the label ${\rm I}$ has been added to denote the heavier radion.
By substituting this expression for $d_2^{{\rm I}'}(y_{\rm I})$ and Eq.\,\eqref{eq:cf2} into Eq.\,\eqref{eq:B_2_yI},
we obtain a quadratic equation for $m^2$.
Then,  for $e^{(2k_2+u_2)y_{\rm I}}/e^{(2k_2+u_2)y_{\rm IR}}\ll 1$, the heavier radion mass-squared is approximately given by
\begin{align}
\label{eq:m_radion_1}
m^2_{r_{\rm I}} \simeq \frac{4}{3}\, l^2 u_1^2 \, \frac{k_2  (2k_1+u_1)}{k_1(k_2-k_1)}\, e^{-2(k_1+u_1)y_{\rm I}} \ .
\end{align}
Note that the mass-squared goes to infinity in the limit of $k_2\to k_1$.
This feature is not an artifact of the approximation but is actually seen in the numerical computation as we will show later. 
For $k_2<k_1$, the mass-squared $m^2_{r_{\rm I}}$ becomes  tachyonic,
and, therefore, we require  $k_2>k_1$.
The same condition was obtained in~\cite{Kogan:2001qx}
where the radion kinetic terms without including the GW field are computed
and it is shown that one radion becomes a ghost for $k_2<k_1$.
This condition is also in accordance with the parameter space that the brane tension on the intermediate brane becomes negative.
The calculation of the same radion mass-squared using the naive ansatz in appendix~\ref{app:naive_ansatz}
also shows the $(k_2 - k_1)$ dependence in the denominator. 
Eq.~\eqref{eq:m_radion_1} is different from that obtained through the naive ansatz
by a factor of about $ e^{-2y_{\rm I}u_1} ({u_1}/{k_1})^{1/2}$.

\begin{figure*}
\begin{minipage}[t]{\hsize}
\includegraphics[width=7cm]{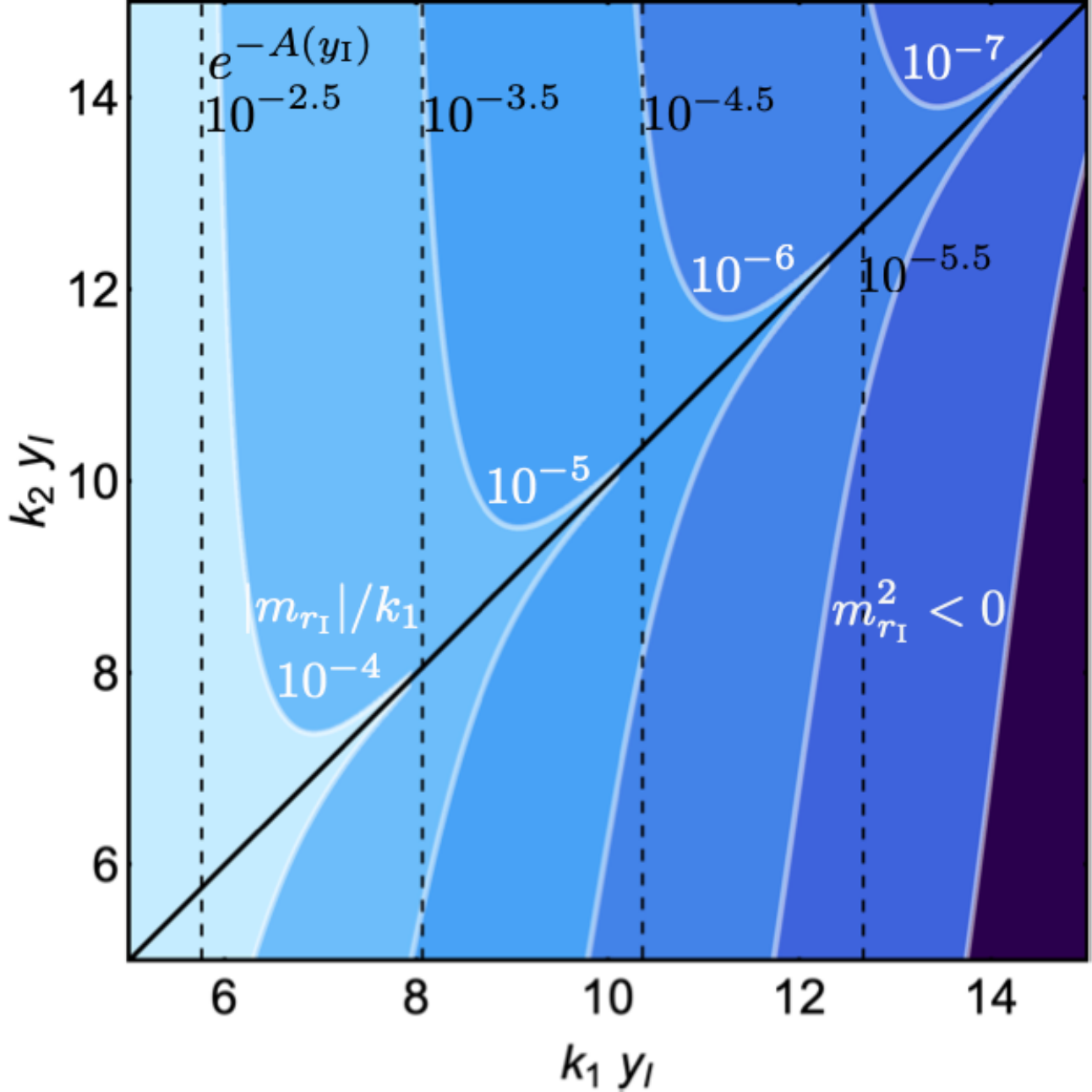}
\hspace{1cm}
\includegraphics[width=7cm]{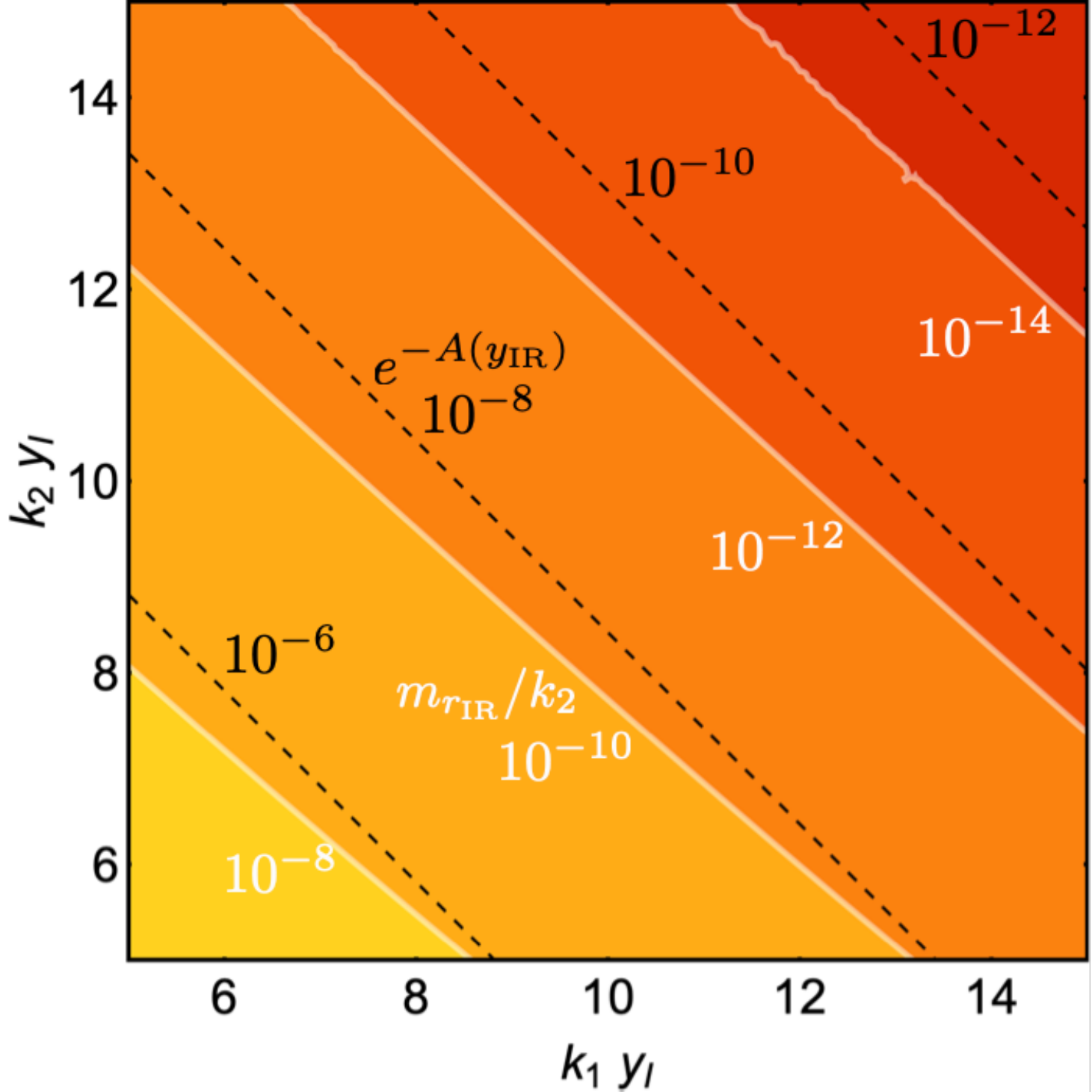}
  \end{minipage}
  \caption{The heavier (left) and lighter (right) radion masses in the three 3-brane model.
  Here, we take $l=0.3,~u_1y_{\rm I}=1.5,~u_2y_{\rm I}=1.5$ and $y_{\rm IR}/y_{\rm I}=2$ as reference values.
  In the left panel, the white solid contours denote the heavier radion mass of 
  $m_{r_{\rm I}}/k_1=10^{-4},~10^{-5},~10^{-6},~10^{-7}$.
  The black dashed lines correspond to the typical mass scale at the intermediate brane of
  $e^{-k_1 y_{\rm I}}=10^{-2.5},~10^{-3.5},~10^{-4.5},~10^{-5.5}$ 
  from the left to the right, respectively.
  The solid black line denotes $k_1=k_2$.
  The radion mass is tachyonic in the region of $k_2<k_1$.
  Besides, $|m_{r_{\rm I}}|$ goes to infinity in the limit of $k_2\to k_1$.
  In the right panel, the white contours represent the lighter radion mass of  
  $m_{r_{\rm IR}}/k_2=10^{-8},~10^{-10},~10^{-12},~10^{-14}$ 
  from the bottom left to the top right.
  The black dashed lines correspond to the typical mass scale at the IR brane of
   $e^{-k_2(y_{\rm IR}-y_{\rm I})-k_1y_{\rm I}}=10^{-6},~10^{-8},~10^{-10},~10^{-12}$ 
  from the bottom left to the top right.}
\label{fig:radion_mass} 
\end{figure*}

Let us now see numerical solutions to the quadratic equation~\eqref{eq:exact_masssq} for $m^2$ 
with the coefficients \eqref{alphabetagamma}.
The left and right panels of Fig.~\ref{fig:radion_mass} show the heavier and lighter radion masses, respectively.
For the heavier radion, the mass becomes tachyonic in the region of $k_2<k_1$,
as obtained via the analytical expression in Eq.\,\eqref{eq:m_radion_1}.
Besides, $|m_{r_{\rm I}}|$ goes to infinity in the limit of $k_2\to k_1$.
Fig.~\ref{fig:radion_mass} also indicates that the heavier (lighter) radion mass is generally more than one order of magnitude
smaller than the typical mass scale on the intermediate (IR) brane,
which is consistent with the approximate analytical expressions.
Such feature has been also seen in ref.~\cite{Agashe:2016rle} in the context of the dual 4D CFT.

For later convenience, we here comment on the ratio of the coefficients $c^{\rm I}_{f1}/c^{\rm I}_{f2}$
for the heavier radion mass solution.
In the approximated expression~\eqref{eq:cf2},
\begin{align}
&c^{\rm I}_{f1}= 
\frac{4\,c_{f2}^{\rm I}
k_1u_1^2(2k_1+u_1)}{-3e^{2(k_1+u_1)y_{\rm I}}k_1(k_2-k_1) m^2
+4k_2u_1^2(2
k_1 +
u_1)}\ ,
\end{align}
we can check that the denominator of the right-hand side is zero
by using the approximated mass-squared formula \eqref{eq:m_radion_1}.
The finite result is obtained by taking account of the terms neglected to get the simple formula. 
In the next section, we evaluate $c^{\rm I}_{f1}/c^{\rm I}_{f2}$ by including the subdominant terms 
and discuss radion couplings to brane-localized matter fields.

%#######################
\section{The radion effective action}
\label{sec:radioneffaction}

We here discuss radion couplings to matter fields localized on branes.
Firstly, the kinetic terms for two radions in the three 3-brane model are found.
Then, the radion-matter couplings are computed in the basis that the radion kinetic terms are canonically normalized.
It will be shown that the lighter (heavier) radion has a dominant coupling to matter fields on the  IR (intermediate) brane
compared to that of the heavier (lighter) radion.

As in the case of the two 3-brane model \cite{Csaki:2000zn}, the kinetic terms of the two radions in the three 3-brane model are calculated 
through the Ricci scalar and bulk cosmological constant terms in the action \eqref{eq:action},
\begin{align}
\label{eq:action_non_o}
&S_r = \sum_{i={\rm I},{\rm IR}}\left[\, -2M_5^3 \int d^4 xdy \sqrt{g(F_i,G_i,E_i)} \, R(F_i,G_i,E_i)
-\int d^4 xdy \sqrt{g(F_i,G_i,E_i)} \, \Lambda^5 \right. \nonumber \\
&\qquad \qquad \quad \left.- \sum_{p={\rm UV},{\rm I},{\rm IR}}\int d^4 x \sqrt{|g^{in}_p(F_i,G_i,E_i)|} \, \chi_i\right.\\
&\qquad \qquad \quad \left.+ \sum_{p={\rm I},{\rm IR}}\int d^4 x \sqrt{|g^{in}_p(F_i,G_i,E_i)|} \, g^{in\,\mu\nu}_p(F_i,G_i,E_i) \, T_{\mu\nu}(x,y_p) \, \right] \nonumber,
\end{align}
where the dependence on the fluctuations is written explicitly.
The label $i={\rm I},{\rm IR}$ denotes the solutions for the heavier and lighter radions in the mass eigenstate basis, respectively,
and the summation is taken over for the two radion solutions.
Note that they are not orthogonal to each other.
The hermiticity condition~\eqref{eq:hermiticity} is not satisfied for the radion mass eigenstates because both of them live in the whole bulk
and $E'$ is nonzero on the intermediate brane.
Still, the action \eqref{eq:action_non_o} leads to two independent equations of motion for the fluctuations labeled by $i={\rm I},{\rm IR}$,
each of which is consistent with the equation of motion \eqref{eq:eq_1}.
The radion kinetic terms also arise from the kinetic term of the GW scalar field,
but their contributions are subdominant due to the suppression of $l^2$.

Considering Eq.~\eqref{eq:F_redefine} with Eqs.~\eqref{f1expression}, \eqref{f2expression},
we can express the two solutions for the fluctuation $F$ as
\begin{align}
&F_{\rm I} (x,y)=\left\{
    \begin{array}{c}
    \left( c^{\rm I}_{f1}e^{2k_1 y} +\frac{A_1'}{2e^{2A_1}}E_{1{\rm I}}' \right) r_{\rm I}(x)\quad (\text{subregion~$1$}) \\[1ex]
    \left( c^{\rm I}_{f2}e^{2k_2 y+2y_{\rm I}(k_1-k_2)} +\frac{A_2'}{2e^{2A_2}}E_{2{\rm I}}' \right) r_{\rm I}(x)\quad (\text{subregion~$2$}) \\
    \end{array}
    \right.  ,\\[2ex]
 &F_{\rm IR} (x,y) =\left\{
    \begin{array}{c}
    \left( c^{\rm IR}_{f1}e^{2k_1 y} +\frac{A_1'}{2e^{2A_1}}E_{1{\rm IR}}' \right) r_{\rm IR}(x)\quad (\text{subregion~$1$}) \\[1ex]
    \left( c^{\rm IR}_{f2}e^{2k_2 y+2y_{\rm I}(k_1-k_2)} +\frac{A_2'}{2e^{2A_2}}E_{2{\rm IR}}' \right) r_{\rm IR}(x)\quad (\text{subregion~$2$}) \\
    \end{array}
    \right. .
\end{align}
Here, $r_{\rm I, IR}(x)$ denote the heavier and lighter radion fields, respectively,
and $c^{\rm I,IR}_{f1}/c^{\rm I,IR}_{f2}$ are given via Eq.\,\eqref{eq:cR_cL}.
We only consider the terms at the zero-th order of $l^2$
because the terms of $\mathcal{O}(l^2)$ give subdominant contributions to the radion kinetic terms. 
By substituting these solutions into the action \eqref{eq:action_non_o} and integrating over $y$,
we find
\begin{align}
\label{eq:radion_kin}
-S &\supset 4M_5^3\int d^4x\, r_{\rm I}\Box r_{\rm I} \, (c_{f2}^{{\rm I}})^2
\biggl[\, \frac{3 (c^{{\rm I}})^2(-1+e^{2k_1 y_{\rm I}})}{k_1} +\frac{3 e^{2k_1 y_{\rm I}}(-1+e^{2k_2(y_{\rm IR}-y_{\rm I})})}{k_2}\nonumber \\[1ex]
&\qquad \qquad  \qquad \qquad \qquad \qquad + \frac{6(c^{{\rm I}}-1)^2\,e^{2k_1y_{\rm I}}}{k_1-k_2} \biggl]  \nonumber \\[1ex]
&\qquad \qquad  +r_{\rm IR}\Box r_{\rm IR} \, (c_{f2}^{{\rm IR}})^2
\biggl[ \, \frac{3 (c^{{\rm IR}})^2(-1+e^{2k_1 y_{\rm I}})}{k_1}+\frac{3 e^{2k_1 y_{\rm I}}(-1+e^{2k_2(y_{\rm IR}-y_{\rm I})})}{k_2}\nonumber \\[1ex]
&\qquad \qquad  \qquad \qquad \qquad \qquad + \frac{6(c^{{\rm IR}}-1)^2\,e^{2k_1y_{\rm I}}}{k_1-k_2}\biggl] \nonumber \\[1ex]
&\equiv \,\,\sum_i\int d^4 x \, M_{\rm pl}^2 \, \frac{1}{2} (\partial r_i)^2 (c_{f2}^{i})^2 K_i^2
= \,\,\sum_i\int d^4 x \, \frac{1}{2} (\partial \tilde r_i)^2 \ ,
\end{align}
where $c^i\equiv c^i_{f1}/c^i_{f2}$, $M_{\rm pl}^2\approx M_5^3/k_1$ denotes the 4D Planck scale
and $K_i$ are dimensionless coefficients.
Eq.\,\eqref{eq:G} has been used to express $G_i$ in terms of $F_i$ and $E_i$.
In the last equality, we have defined canonically normalized radion fields $\tilde r_i(x)\equiv M_{\rm pl}c^i_{f2}K_i r_i(x)$.
Note that the bulk profiles of the fluctuations $E_i$ are not necessary to obtain Eq.~\eqref{eq:radion_kin}.
For the kinetic term of the radion $r_{\rm I}$, the first and third terms are dominant,
and it is approximated as
\begin{align}
4M_5^3\int d^4x\, r_{\rm I}\Box r_{\rm I} \, (c_{f2}^{{\rm I}})^2
\biggl[\, \frac{3 (c^{{\rm I}})^2(e^{2k_1 y_{\rm I}})(k_1+k_2)}{k_1(k_1-k_2)} \biggl]  \ .
\end{align}
This expression indicates that the $r_{\rm I}$ kinetic term becomes negative for $k_2<k_1$ where the radion is a ghost.
This condition is the same as that leading to the tachyonic radion mass in Eq.~\eqref{eq:m_radion_1}.

\begin{table}[!t]
\caption{The radion couplings to matter fields living on the intermediate and IR branes.
Two example cases are considered as sample A and sample B.
The left table summarizes input parameters in unit of $M_5$.
In the right table, $\tilde m_{r_{\rm I}}$ and $\tilde m_{r_{\rm IR}}$ are, respectively, the normalized heavier and lighter radion masses defined as
$\tilde m_{r_{\rm I}} \equiv m_{r_{\rm I}}(k_1e^{-k_1y_{\rm I}})^{-1}$ and 
$\tilde m_{r_{\rm IR}} \equiv m_{r_{\rm IR}}(k_2e^{-k_2(y_{\rm IR}-y_{\rm I})-k_1y_{\rm I}})^{-1}$.
The normalized radion couplings to matter fields on the intermediate and IR brane
$\tilde C_{y_{\rm I}}(m_{r_{\rm I}})$, $\tilde C_{y_{\rm IR}}(m_{r_{\rm I}})$, $\tilde C_{y_{\rm I}}(m_{r_{\rm IR}})$,
$\tilde C_{y_{\rm IR}}(m_{r_{\rm IR}})$ are  defined as
$C_{y_j}(m_{r_{i}}) \equiv \frac{1}{M_{\rm pl}e^{-A(y_j)}}\tilde C_{y_j}(m_{r_{i}})$
where the factor $M_{\rm pl}\,e^{-A(y_{j})}$ in the denominator denotes the typical mass scale of the brane at $y=y_j$.
The results in the table indicate that the heavier (lighter) radion dominantly couples to matter fields on the intermediate (IR) brane.}
\label{tab:radion-matter}
\begin{center}
\begin{tabular}{lrr} 
\toprule
\multicolumn{3}{c}{{\bf Input}}\\
 & {Sample A} & { Sample B} \\
\cmidrule(r){2-2}
\cmidrule(l){3-3}
$y_{\rm I}$ & $10$  & $10$   \\
$y_{\rm IR}$ & $15$  &  $20$  \\
$k_1$ & $0.3$  & $0.3$   \\
$k_2$ & $0.6$ & $0.6$   \\
$u_1$ & $0.05$  & $0.05$    \\
$u_2$ & $0.05$   & $0.05$    \\
\bottomrule
\end{tabular}
~~~~~
\begin{tabular}{lrr}
\toprule
\multicolumn{3}{c}{{\bf Output}}\\
 & {Sample A} & { Sample B} \\
\cmidrule(r){2-2}%
\cmidrule(l){3-3}
$\tilde m_{r_{\rm I}}$ & $0.24$  & $0.24$   \\
$\tilde C_{y_{\rm I}}( m_{r_{\rm I}})$ & $-0.47$  & $-0.47$   \\
$\tilde C_{y_{\rm IR}}( m_{r_{\rm I}})$ & $1.7\times 10^{-5}$ & $1.27\times 10^{-9}$   \\
$\tilde m_{r_{\rm IR}}$ & $0.066$  &  $0.051$  \\
$\tilde C_{y_{\rm I}}( m_{r_{\rm IR}})$ & $5.7\times 10^{-5}$  & $1.8\times 10^{-6}$    \\
$\tilde C_{y_{\rm IR}}( m_{r_{\rm IR}})$ & $0.58$  & $0.58$    \\
\bottomrule
\end{tabular}
\end{center}
\end{table}

The radion-matter interaction terms are written for matter fields living on the intermediate and IR branes, respectively, as
\begin{align}
&-\sqrt{g^{in}}g^{in\,\mu\nu}T_{\mu\nu}|_{y=y_{\rm I}} =2F_{1i}(x,y_{\rm I}) \hat T(x,y_{\rm I}) 
\equiv C_{y_{\rm I}}(m_{r_i})\,\tilde r_i(x)\hat  T(x,y_{\rm I}) \, , \\[1ex]
&-\sqrt{g^{in}}g^{in\,\mu\nu}T_{\mu\nu}|_{y=y_{\rm IR}} =2F_{2i}(x,y_{\rm  IR}) \hat  T(x,y_{\rm IR})
\equiv C_{y_{\rm IR}}(m_{r_i})\,\tilde r_i(x)\hat  T(x,y_{\rm IR})
\ ,
\end{align}
where $i={\rm I},{\rm IR}$ and $\hat  T(x,y_p)$ is the normalized trace of the energy momentum tensor of matter fields on a brane, $i.e.$ $\hat  T=e^{-2A}\eta^{\mu\nu}T_{\mu\nu}$.%
\footnote{We use the normalized trace of $T_{\mu\nu}$ to consider canonically normalized fields in $\hat T$.
For example, a real scalar field $\chi$ on a brane at $y=y_i$ is included as $T_{\mu\nu}\supset \frac{1}{2}\partial_\mu\chi\partial_\nu\chi$.
Then, the action is $S\supset \int d^4x \sqrt{g(y_i)}g^{\mu\nu}(y_i)T_{\mu\nu}\supset \int d^4x \,e^{-2A(y_i)} \eta^{\mu\nu}(1-2F(y_i))\frac{1}{2}\partial_\mu\chi\partial_\nu\chi$. Thus, the canonically normalized field $\tilde \chi$ is obtained by $\chi\equiv e^{A} \tilde  \chi $.}
The constant $C_{y_j}(m_{r_i})$ denotes the radion-matter coupling at $y=y_j$ for the radion with mass $m_{r_i}$.
In the basis that the radion kinetic terms are canonically normalized, these couplings are given by
\begin{align}
\label{eq:radion_matter_yI_cn}
C_{y_{\rm I}}(m_{r_i}) &=  \frac{2}{K_i M_{\rm pl}} \frac{c^i A_2'-A_1'}{A_2'-A_1'} \, e^{2A_1(y_{\rm I})}  \, , \\[1ex]
\label{eq:radion_matter_yIR_cn}
C_{y_{\rm IR}}(m_{r_i}) &= \frac{2}{K_i M_{\rm pl}} \, e^{2A_2(y_{\rm IR})} \ .
\end{align}
Tab.~\ref{tab:radion-matter} shows the radion couplings to matter fields living on the intermediate and IR branes for some example parameters.
We can see that
a radion couples to matter fields on a brane
as $\sim \frac{1}{M_{\rm pl}\,e^{-A(y_{j})}}\tilde r_i(x) \hat{T}(x,y_{j})$
where the factor $M_{\rm pl}\,e^{-A(y_{j})}$ in the denominator denotes the typical mass scale of the brane at $y=y_j$.
Furthermore, the lighter radion coupling to matter fields on the intermediate brane is suppressed compared to that on the IR brane
as the lighter radion is localized toward the IR brane.
The suppression of the lighter radion coupling to matter fields on the intermediate brane has been also discussed in ref.~\cite{Agashe:2016rle} through the dual CFT picture.
In our analysis, a further suppression could be seen when $e.g.$ $(y_{\rm IR}-y_{\rm I})/y_{\rm I}\ll 1$ is not satisfied.

%#######################
\section{Discussions}\label{discussions}

We have presented the stabilization of two radions in warped extra dimension models with multiple 3-branes
through a simple extension of the GW mechanism by introducing a single 5D scalar field.
The metric for each bulk subregion bounded by two 3-branes has a different warp factor in general.
We have mainly focused on the three 3-brane set-up for simplicity, but
it was straightforward to generalize the discussion to models with any number of 3-branes as shown in Appendix~\ref{app:generalN}.
The stabilization of radions was shown by solving the Einstein equation and the scalar field equation of motion simultaneously
so that the backreaction effect from the GW scalar field background is taken into account.
We then considered perturbations from the background configuration and
computed the mass spectrum of the radion-scalar field system and also the spectrum of KK gravitons by solving the equations of motion.
Nonzero masses of radions were found by taking account of the backreaction effect of the GW scalar field background.
The lighter radion has a mass close to the typical mass scale of the IR brane
while the heavier radion has a mass at the order of the scale of the intermediate brane. 
The radion couplings to brane-localized matter fields have been also calculated. 
It was shown that the lighter (heavier) radion has a dominant coupling to matter fields on the IR (intermediate) brane.

\begin{figure}[!t]
    \includegraphics[width=2.5in]{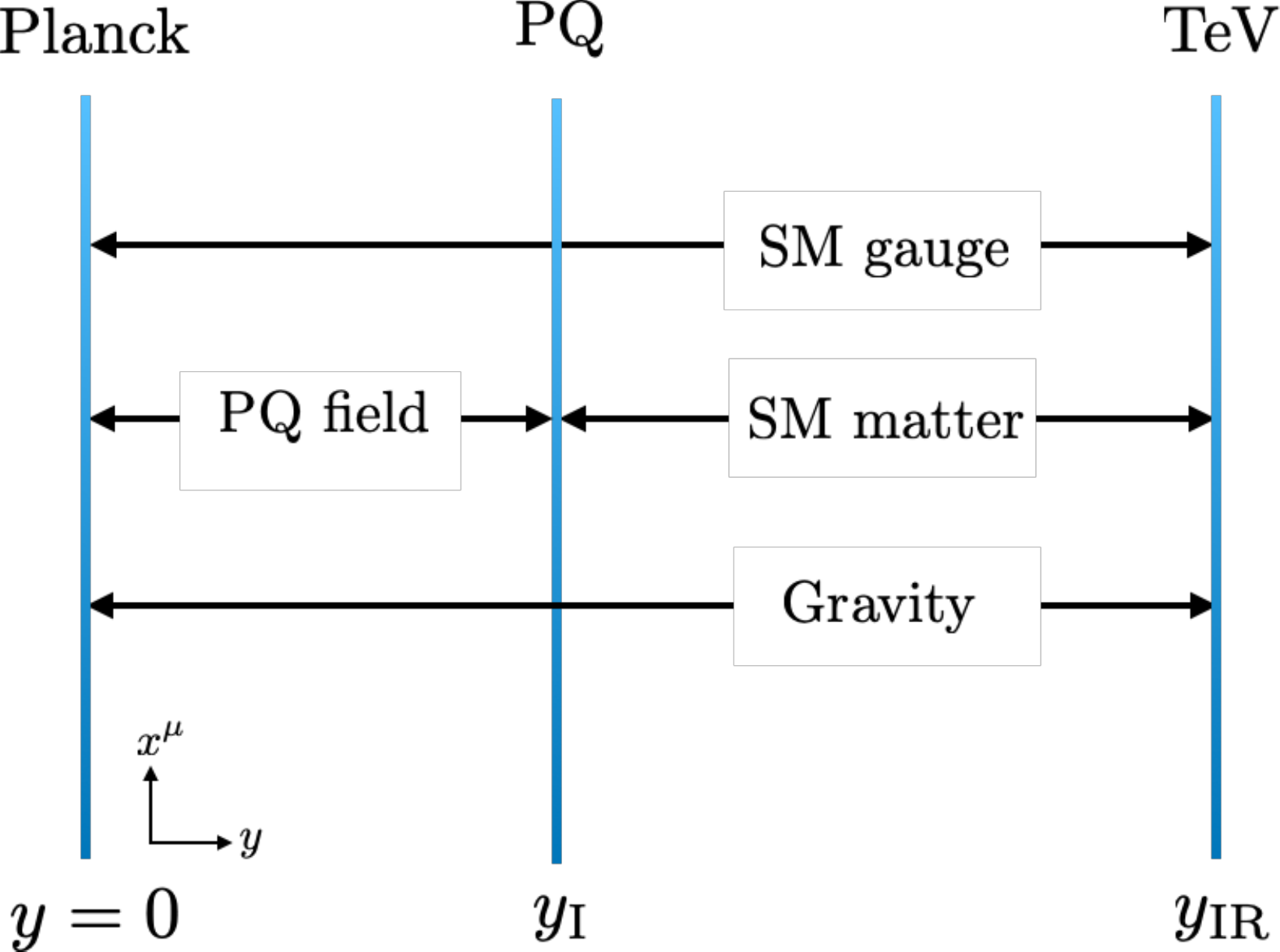}
    \vspace{0.5cm}
  \caption{The three 3-brane model to realize a natural and high-quality axion solution to the strong CP problem.
  The typical mass scales of the UV, intermediate and IR branes are identified as the Planck, PQ breaking and TeV scales.
  The $U(1)_{\rm PQ}$ gauge field and a PQ breaking scalar field live in the bulk subregion between the Planck and PQ branes
  while the SM fields live in the other subregion between the PQ and TeV branes.
  A pair of KSVZ (anti-)quarks charged under the $U(1)_{\rm PQ}$ live on the intermediate brane. 
  Gravity propagates in the whole bulk spacetime.}
  \label{fig:applications1}
\end{figure}

Warped extra dimension models with multiple 3-branes have a huge potential for innovation in model-building of physics beyond the SM
which often indicates the existence of new mass scales hierarchically different from both the Planck scale and the electroweak scale.
Since the typical energy scale of each brane is exponentially different from those of the other branes,
multiple 3-brane models enable us to explain multiple hierarchical mass scales naturally.
For instance, the PQ solution to the strong CP problem
introduces an intermediate energy scale
where the $U(1)_{\rm PQ}$ symmetry is spontaneously broken.
Without any mechanism, this new energy scale leads to another naturalness problem as in the case of the SM Higgs field:
the $U(1)_{\rm PQ}$ breaking scale is unstable under radiative corrections and its smallness compared to the Planck scale requires a fine-tuning.
Furthermore, the usual PQ solution suffers from the so-called axion quality problem.
The $U(1)_{\rm PQ}$ global symmetry must be realized to an extraordinarily high degree while
it has been discussed that quantum gravity effects do not respect such a global symmetry.
Our three 3-brane model can address these problems as well as the electroweak naturalness problem simultaneously.
As described in Fig.~\ref{fig:applications1},
we can identify the typical mass scales of the UV, intermediate and IR branes as the Planck, PQ breaking and TeV scales.
A scalar field, which spontaneously breaks the $U(1)_{\rm PQ}$ (gauge) symmetry,
lives in the bulk subregion between the Planck and PQ branes
while the SM fields live in the other subregion between the PQ and TeV branes.
Following the two 3-brane model discussed in ref.~\cite{Cox:2019rro} (see also ref.~\cite{Nakai:2021nyf} for a dual 4D realization with SUSY),
a PQ brane-localized potential drives a nonzero VEV for the PQ breaking scalar field localized toward the PQ brane.
In the same way as the original RS model for the electroweak breaking, the naturalness problem for the PQ breaking scale can be addressed.
Since the bulk $U(1)_{\rm PQ}$ gauge symmetry is broken at the UV brane,
we generally expect $U(1)_{\rm PQ}$-violating Planck suppressed operators.
However, the scalar field profile is exponentially small at the UV brane, and hence
these dangerous operators are suppressed enough to achieve the high-quality axion.
The correct axion-gluon coupling to solve the strong CP problem is introduced
by for example a pair of (anti-)quarks charged under the $U(1)_{\rm PQ}$
which live on the PQ brane in the same way as the so-called KSVZ axion model
\cite{Kim:1979if,Shifman:1979if}.
The SM Higgs field lives on (or is localized toward) the TeV brane so that the electroweak naturalness problem is addressed as usual.

\begin{figure}[!t]
    \includegraphics[width=2.5in]{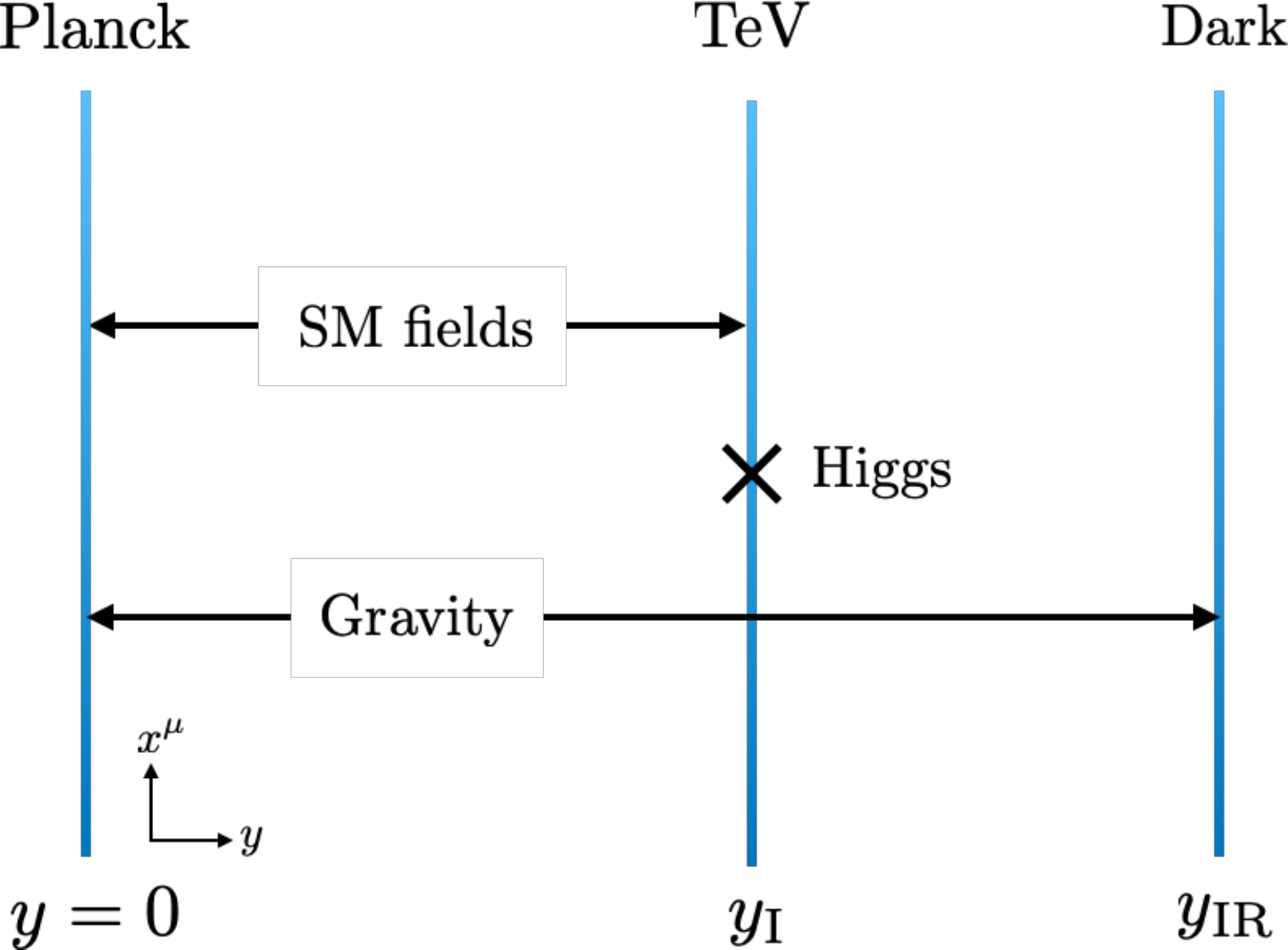}
    \vspace{0.5cm}
  \caption{The three 3-brane model to introduce a dark sector.
  The typical mass scales of the UV, intermediate and IR branes are identified as the Planck, TeV and some smaller dark sector mass scales.
Here, we call the IR brane as the dark brane.
The SM fields live in the bulk subregion between the Planck and TeV branes, while the SM Higgs field is localized at the TeV brane.
Gravity propagates in the whole bulk spacetime.
}
  \label{fig:applications2}
\end{figure}

Another interesting possibility in the three 3-brane model framework is to identify the typical mass scales of the UV, intermediate and IR branes as
the Planck, TeV and some smaller scales.
We call the IR brane in this case as the {\it dark brane}.
As described in Fig.~\ref{fig:applications2}, the SM fields live in the bulk subregion between the Planck and TeV branes.
The SM Higgs field is localized at the TeV brane so that the electroweak naturalness problem is addressed.
Only gravity (and possibly some new light particles feebly interacting with the SM fields) can live in the subregion between the TeV and dark branes
\cite{Agashe:2016rle, Cai:2021nmk}
(see also ref.~\cite{Georgi:2016xhm} for a dual 4D description where the lighter radion is roughly identified as the lightest scalar glueball).
The mass scale of the dark brane can be exponentially smaller than the TeV scale.
In this case, since the lighter radion coupling to the SM fields is suppressed,
its lifetime is long enough to become a dark matter candidate.

The RS model has been known to predict a characteristic cosmological history.
The system at a high temperature is described by the de-compactified AdS-Schwarzschild (AdS-S) spacetime where the IR brane is replaced by an event horizon
\cite{Creminelli:2001th}.
Then, as the Universe cools down, a phase transition from the AdS-S spacetime to the compactified RS spacetime takes place.
In the dual 4D picture, it can be understood as a confinement-deconfinement phase transition.
The phase transition is typically of the strong first order and experiences a supercooling phase before nucleation of true vacuum bubbles develops.
During such a transition, detectable gravitational waves (GWs) are generated.
The cosmological history of multiple 3-brane models is even more fascinating.
As the Universe cools down, multiple confinement-deconfinement phase transitions take place.
Correspondingly, sizable GWs with multiple peak frequencies are produced.
The three 3-brane setup in Fig.~\ref{fig:applications1} predicts a first-order phase transition at the temperature of the PQ breaking scale,
which generates GWs detectable at LIGO~\cite{LIGOScientific:2014pky} (see ref.~\cite{DelleRose:2019pgi} for the case of the two 3-brane model),
as well as a first-order transition at the temperature of the electroweak scale
which produces GWs to be observed at future space-based GW observers
such as TianQin~\cite{TianQin:2015yph}, Taiji~\cite{Hu:2017mde}, eLISA~\cite{eLISA:2013xep}, DECIGO~\cite{Seto:2001qf}
and BBO~\cite{Harry:2006fi}.
On the other hand, the three 3-brane model in Fig.~\ref{fig:applications2} predicts
a first-order phase transition at the temperature of the electroweak scale and also
a first-order transition at some low temperature which leads to low frequency GWs.
Interestingly, pulsar timing data recently reported by the NANOGrav collaboration may indicate the existence of
a stochastic GW background around $f \sim 10^{-8}$ Hz
\cite{NANOGrav:2020bcs}.
It has been pointed out in ref.~\cite{Nakai:2020oit} that the reported signal can be interpreted as GWs from a strong first-order phase transition
at which the temperature of the SM sector is $\mathcal{O}(1-100) \, \rm MeV$.
Such a phase transition can be a natural consequence of the three 3-brane model.

%#######################
\section*{Acknowledgements}

We would like to thank Sungwoo Hong, Ryo Namba and Tsutomu Yanagida for useful discussions.
S.L.\ was supported by the National Research Foundation of Korea (NRF) grant funded by the Korea government (MEST) (No. NRF-2021R1A2C1005615).
S.L.\ was also supported by the Visiting Professorship at Korea Institute for Advanced Study.

\appendix

\section{Radion dynamics via the naive ansatz}
\label{app:naive_ansatz}

In this appendix, we discuss the stabilization of two radions for the three 3-brane system
in a similar way as the original discussion of the GW mechanism~\cite{Goldberger:1999uk,Goldberger:1999un}.
In the main text, the brane separation distances in Eqs.\,\eqref{eq:GW_y_I} and \eqref{eq:GW_y_IR} were obtained
by solving the Einstein equation and the GW field equation of motion simultaneously.
Here, we will solve a bulk equation for the GW scalar field in a background metric without the backreaction effect.
By inserting the solution to the action and integrating over the extra dimension,
the effective potential for the brane separation distances is obtained.
The distances are then fixed at the minimum of the potential.
We will also compute the masses of radions.
In the main text, they were found by solving the equations of spin-0 fluctuations
including the backreaction effect from the GW scalar field background.
Instead, we take a naive ansatz for the metric
and compute the radion masses through the effective potential.
The condition to avoid tachyonic masses of radions is obtained from the kinetic terms.

%#######################
\subsection{Stabilization of distances}
\label{sec:radion_stabilization}

Let us consider a bulk real scalar field $\Phi$ as the GW field whose action is given by%
\footnote{We here use a notation for the GW field different from that of the main text.}
\begin{align}
\label{naiveaction}
S=\,\, &\frac{1}{2}\int d^4 x \int dy \sqrt{g}\left(g^{MN}\partial_M\Phi \partial_N\Phi-m^2_\Phi \Phi^2\right) \nonumber \\
&-\int d^4 x \int dy \sqrt{g} \, \Bigl[\, \lambda_{\rm UV} (\Phi^2-v_{\rm UV}^2)^2\,\delta(y) \nonumber \\
&\qquad \qquad \qquad \qquad +\lambda_{\rm I}(\Phi^2-v_{\rm I}^2)^2\,\delta(y-y_{\rm I})
+\lambda_{\rm IR}(\Phi^2-v_{\rm IR}^2)^2\,\delta(y-y_{\rm IR}) \, \Bigr]\ ,
\end{align}
where $m_\Phi$ denotes a mass parameter for the GW field, $\lambda_p$ ($p=\rm UV, I, IR$) are dimensionless couplings
and $v_p$ are parameters with mass dimension $3/2$.
The mass parameter $m_{\Phi}$ may be different for the subregion $1,2$ as in the case of the main text,
and we label them as $m_{\Phi_{1,2}}$, respectively.
The terms in the second and third lines represent brane-localized potentials.
We use the metric in Eq.\,\eqref{metric} without including the backreaction effect from the GW field.
The bulk equation of motion for $\Phi$ is
\begin{align}
\Phi''-4A'\Phi'-m_\Phi^2\Phi
=0\ .
\end{align}
Then, we obtain the following general solution in the bulk, 
\begin{align}
\label{generalsol}
\Phi (y)=e^{2A' y}\left[ \, \tilde A e^{\nu A' y}+\tilde Be^{-\nu A' y} \, \right] .
\end{align}
Here, the constants $(\tilde A,\tilde B,\nu) \equiv (\tilde A_1,\tilde B_1,\nu_1)$ and $(\tilde A_2,\tilde B_2,\nu_2)$ for $0< y < y_{\rm I}$
and $y_{\rm I}<y < y_{\rm IR}$, respectively,
and we have defined $\nu_1^2 \equiv 4+m_{\Phi_1}^2/{A'}^2=4+m_{\Phi_1}^2/k_1^2$ and
$\nu_2^2 \equiv 4+m_{\Phi_2}^2/{A'}^2 =4+m_{\Phi_2}^2/k_2^2$.
Inserting the solution into the action \eqref{naiveaction} and integrating over $y$, we find the effective potential,
\begin{align}
\label{effpotential}
V(y_{\rm I},y_{\rm IR})=\,\,
&\left(-e^{-2k_1y_{\rm I}\nu_1}+1\right)k_1 \left\{\tilde A_1^2e^{2k_1y_{\rm I}\,\nu_1}(2+\nu_1)+\tilde B_2^2(\nu_1-2)\right\} \nonumber \\
&-\left(e^{4y_{\rm I}(k_2-k_1)-2k_2y_{\rm IR}\nu_2}-e^{4y_{\rm I}(k_2-k_1)-2k_2y_{\rm I}\nu_2}\right)
\nonumber \\&\qquad \times
k_2\left\{\tilde A_2^2e^{2k_2(y_{\rm I}+y_{\rm IR})\nu_2}(\nu_2+2)+\tilde B_2^2(\nu_2-2)\right\} \nonumber \\
&+\lambda_{\rm UV} \left(\Phi^2(0)-v_{\rm UV}^2 \right)^2+\lambda_{\rm I} \left(\Phi^2(y_{\rm I})-v_{\rm I}^2 \right)^2
+\lambda_{\rm IR} \left(\Phi^2(y_{\rm IR})-v_{\rm IR}^2 \right)^2  .
\end{align}
By taking the limit $\lambda_{\rm UV,I,IR}\to \infty$, we obtain the Dirichlet boundary conditions,
\begin{align}
\Phi(0)=v_{\rm UV} \, , \qquad \Phi(y_{\rm I})=v_{\rm I} \, , \qquad \Phi(y_{\rm IR})=v_{\rm IR}\ .
\end{align}
The general solution \eqref{generalsol} must satisfy these boundary conditions.
Note that the condition at the intermediate brane is imposed on the solutions for both subregions.
Then, $\tilde A_{1,2},\tilde B_{1,2}$ are determined as
\begin{align}
&\tilde A_1=\frac{X^{2+\nu_1}(-v_{\rm I}+v_{\rm UV} X^{\nu_1-2})}{-1+X^{2\nu_1}} \ ,
\quad \tilde B_1=\frac{-v_{\rm UV}+v_{\rm I}X^{2+\nu_1}}{-1+X^{2\nu_1}} \ ,\\[1ex]
&\tilde A_2=\frac{Y^{2+\nu_2}(v_{\rm IR} -v_{\rm I}X^{k_2/k_1(2-\nu_2)}Y^{\nu_2-2})}{1-Y^{2\nu_2}/X^{2k_2\nu_2/k_1}} \ ,
\quad \tilde B_2=\frac{-v_{\rm I}X^{2k_2/k_1+k_2\nu_2/k_1}+v_{\rm IR}Y^{2+\nu_2}}{-X^{2k_2\nu_2/k_1}+Y^{2\nu_2}} \ ,
\end{align}
where we have defined $X \equiv e^{-k_1y_{\rm I}}$ and $Y \equiv e^{-k_2y_{\rm IR}}$.
Inserting these constants into the effective potential \eqref{effpotential} for the brane separation distances,
we find
\begin{align}
\label{eq:radion_potential}
\begin{split}
&V=V_1+V_2\ ,\\[1ex]
&V_1=4k_1X^4(v_{\rm I}-v_{\rm UV}X^{\epsilon_1})^2\left(1+\frac{\epsilon_1}{4}\right)+\epsilon_1k_1v_{\rm UV}^2-\epsilon_1k_1v_{\rm UV}X^{4+\epsilon_1}(2v_{\rm I}-v_{\rm UV}X^{\epsilon_1})\  ,\\[1ex]
&V_2=4k_2Z^4 \left\{v_{\rm IR}-v_{\rm I} \left(\frac{Z}{X} \right)^{\epsilon_2} \right\}^2\left(1+\frac{\epsilon_2}{4}\right) \\
&\qquad +\epsilon_2 k_2v_{\rm I}^2X^{4}
-\epsilon_2k_2v_{\rm I}Z^{4}\left(\frac{Z}{X} \right)^{\epsilon_2} \left\{2v_{\rm IR}-v_{\rm I}\left(\frac{Z}{X} \right)^{\epsilon_2} \right\} \ ,
\end{split}
\end{align}
with $\epsilon_{1,2} \equiv \nu_{1,2}-2$ and $Z \equiv YX^{k_2/k_1-1}=e^{-k_2(y_{\rm IR}-y_{\rm I})-k_1y_{\rm I}}$.
Here, $X\gg Z$ has been used to reduce the expression, and the terms of $\mathcal{O}(\epsilon^2)$ have been ignored.
Ignoring the terms of $\mathcal{O}(\epsilon)$ in the potential $V_1$ and assuming $0<v_{\rm I}/v_{\rm UV}<1$,
the first term of $V_1$ leads to the minimum at 
\begin{align}
\label{distance1}
X^{\epsilon_1}= \frac{v_{\rm I}}{v_{\rm UV}} \ ,
\end{align} 
which can stabilize the distance $y_{\rm I}$.
Similarly, ignoring the terms of $\mathcal{O}(\epsilon)$ in the potential $V_2$ and assuming $0<v_{\rm IR}/v_{\rm I}<1$,
the first term of $V_2$ gives the minimum at 
\begin{align}
\label{distance2}
\left(\frac{Z}{X} \right)^{\epsilon_2} = \frac{v_{\rm IR}}{v_{\rm I}} \ ,
\end{align}
which can stabilize the distance $(y_{\rm IR}-y_{\rm I})$.
The brane separation distances obtained in Eqs.~\eqref{distance1}, \eqref{distance2} are similar to those found in the main text.
The potential similar to Eq.~\eqref{eq:radion_potential} has been discussed
in the multi-brane setup of refs.~\cite{Agashe:2016kfr,Choudhury:2000wc}.

\subsection{Masses of radions}
\label{app:radion_mass_naive}

We now calculate the radion masses in the three 3-brane model by extending the discussion of the two 3-brane setup. 
In the subregion $0\leq y\leq y_{\rm I}$, we take an ansatz for the metric,
\begin{align}
\label{eq:metric_ansatz_1}
ds^2=e^{-2k_1\phi T_1(x)}\eta_{\mu\nu}dx^\mu dx^\nu-T_1(x)^2 d\phi^2\ ,
\end{align}
with $0\leq \phi\leq \phi_{\rm I}$.
This metric ansatz is the same as that of the two 3-brane setup~\cite{Goldberger:1999un}.
On the other hand, for the subregion $y_{\rm I}\leq y\leq y_{\rm IR}$, we use an ansatz,
\begin{align}
\label{eq:metric_ansatz_2}
ds^2=e^{-2(k_2(\phi'-\phi'_{\rm I}) T_2(x)+k_1\phi_{\rm I} T_1(x))}\eta_{\mu\nu}dx^\mu dx^\nu-T_2(x)^2 d\phi'^2\ ,
\end{align}
with $\phi'_{\rm I}\leq \phi'\leq \phi'_{\rm IR}$.
Then, the action is given by
\begin{align}
\label{eq:action_radion_naive}
\begin{split}
S&=-4M_5^3\int d^4x\int_0^{\phi_{\rm I}} d\phi \sqrt{g} \, e^{-4k_1\phi T_1(x)} \, T_1(x) R\\
&\quad -4M_5^3\int d^4x\int_{\phi'_{\rm I}}^{\phi'_{\rm IR}} d\phi' \sqrt{g}
\, e^{-4(k_2(\phi'-\phi'_{\rm I}) T_2(x)+k_1\phi_{\rm I} T_1(x))} \, T_2(x) R \\[1ex]
&=24M_5^3\left(\frac{1}{k_1}-\frac{1}{k_2}\right) \int d^4x\sqrt{g} \,
\frac{1}{2}\eta^{\mu\nu} \bigl(\partial_\mu e^{-k_1\phi_{\rm I}T_1} \bigr) \bigl(\partial_\nu e^{-2k_1\phi_{\rm I}T_1} \bigr)  \\[1ex]
&\quad +\frac{24M_5^3}{k_2} \int d^4x\sqrt{g} \, \frac{1}{2}\eta^{\mu\nu} \bigl(\partial_\mu e^{-k_2(\phi_{\rm UV}'-\phi_{\rm I}') T_2(x)-k_1\phi_{\rm I} T_1(x)} \bigr)  \\
&\qquad \qquad \qquad \qquad \qquad \qquad  \times \bigl(\partial_\nu e^{-k_2(\phi_{\rm UV}'-\phi_{\rm I}') T_2(x)-k_1\phi_{\rm I} T_1(x)} \bigr)\ .
\end{split}
\end{align}
For one radion degree of freedom to have the correct kinetic term, we require
\begin{align}
k_2>k_1\ .
\end{align}
This condition is consistent with that obtained in ref.~\cite{Kogan:2001qx} or the condition to avoid a tachyoic radion mass
presented in the main text (see Sec.\,\ref{sec:radion_mass}).
We redefine $T_{1,2}(x)$ as
\begin{align}
\varphi_1&\equiv f_1e^{-k_1\phi_{\rm I} T_1}\ , \quad f_1=\sqrt{24M_5^3\left(\frac{1}{k_1}-\frac{1}{k_2}\right)}\ ,\\[1ex]
\varphi_2&\equiv f_2e^{-k_2(\phi_{\rm UV}'-\phi_{\rm I}') T_2(x)-k_1\phi_{\rm I} T_1(x)}\ , \quad f_2=\sqrt{\frac{24M_5^3}{k_2}}\ ,
\end{align}
Eq.\,\eqref{eq:action_radion_naive} is then rewritten as
\begin{align}
S&=\int d^4x\sqrt{g} \, \left[\, \frac{1}{2} \eta^{\mu\nu}\partial_\mu\varphi_1\partial_\nu\varphi_1+ \frac{1}{2}\eta^{\mu\nu}\partial_\mu\varphi_2\partial_\nu\varphi_2 \, \right] \  .
\end{align}
As the canonically normalized radion fields $\varphi_{1,2}$ have been found,
their masses $m_{\varphi_{1,2}}$ are calculated through the potential \eqref{eq:radion_potential}.
That is, we identify $X$ as $\varphi/f_1$ in $V_1$ and compute $\partial_{\varphi_1}^2V_1$ around the potential minimum
to find the mass $m_{\varphi_{1}}$ in the same manner as the case of the two 3-brane model~\cite{Csaki:1999mp}.
The other radion mass $m_{\varphi_2}$ is also obtained via the potential $V_2$ by identifying $Z$ as $\varphi_2/f_2$
and taking $X = e^{-k_1y_{\rm I}}$.
Then, we find
\begin{align}
m^2_{\varphi_1} & =\frac{2}{3}\epsilon_1^{3/2} \, \frac{v_{\rm I}^2}{M_5^3} \frac{k_1^2 k_2}{k_2-k_1} \,  e^{-2k_1y_{\rm I}}
\sim \frac{16}{3}l^2\left(\frac{u_1}{k_1}\right)^{3/2} \frac{k_1^2 k_2}{k_2-k_1} \, e^{-2A(y_{\rm I})} \ ,
\\[1ex]
m^2_{\varphi_2}
&= \frac{2}{3}\epsilon_2^{3/2} \, \frac{k_2^2v_{\rm IR}^2}{M_5^3}  \, e^{-2(k_2(y_{\rm IR}-y_{\rm I})+k_1y_{\rm I})}
\sim \frac{16}{3}l^2\left(\frac{u_2}{k_2}\right)^{3/2}k_2^2 \, e^{-2A(y_{\rm IR})}
\ ,
\end{align}
at the leading order of $\epsilon_{1,2}$.
The last expressions in both equations are written in terms of the parameters defined in the main text,
identifying them as $m^2_{\Phi_{1,2}}\sim 4u_{1,2}k_{1,2}$,
$v_{\rm UV,I,IR}\sim \phi(y_{\rm UV,I,IR})$ and $\epsilon_{1,2}\approx {(1/4)m_{\Phi_{1,2}}^2}/{k_{1,2}^2}\sim {u_{1,2}}/{k_{1,2}}$.
We have also used ${v_{\rm IR}^2}/{M_5^3} \sim {v_{\rm UV}^2}/{M_5^3}\sim 8l^2$.
The obtained $m_{\varphi_{1,2}}^2$ have the similar orders of magnitude as the results in
Eq.\,\eqref{eq:m_radion_1} and Eq.\,\eqref{eq:m_radion_2}.
However, their scalings of $\epsilon_{1,2}$ are different from Eq.\,\eqref{eq:m_radion_1} and Eq.\,\eqref{eq:m_radion_2}.
The similar differences have been also seen in the case of the two 3-brane model~\cite{Csaki:2000zn}.

\section{Perturbations in the $N+1$ 3-brane system}
\label{app:generalN}

In section~\ref{sec:metric_perturbation}, we have obtained the bulk equation for the fluctuation $f$ in each subregion
for the three 3-brane model as in Eqs.~\eqref{eq:bulk_f_1}, \eqref{eq:bulk_f_2}.
The bulk equations are solved with the boundary conditions \eqref{boundaryUV}, \eqref{boundaryIR}, \eqref{E'continuity}, \eqref{Fcontinuity}.
The discussion can be generalized to the $N+1$ 3-brane system presented in section~\ref{sec:background_metric}.
As described in Fig.~\ref{fig:N_1_branes}, $N+1$ 3-branes are placed at points $y = y_0\,(=0),y_1,y_2, \cdots,y_{N-1},y_N$
where $y_0,y_N$ are the orbifold fixed points.
The bulk equation for the fluctuation $f_p$ defined in the subregion $p$ ($y_{p-1} < y < y_p$) is given by
\begin{align}
&f_p''-f_p'\left(2A_p'+2\frac{\phi_{0,p}''}{\phi_{0,p}'}\right)-f_p\left(4 A_p''-4A_p' \frac{\phi_{0,p}''}{\phi_{0,p}'}\right)=-m^2 e^{2A_p}f_p
\quad \text{(subregion $p$)}\ .
\end{align}
Here, $\Box f_p=-m^2 f_p$ is used.
A set of bulk equations are solved under the boundary conditions at the orbifold fixed points,
\begin{align}
&f_{1}'-2 A_{1}' f_{1}|_{y=0}=0\ , \\[1ex]
&f_{N}'-2 A_{N}' f_{N}|_{y=y_N}=0\ ,
\end{align}
and the conditions at the intermediate branes,
\begin{align}
&f_{q}'-2A_{q}' f_{q}+\frac{f_q-f_{q+1}}{A_{q+1}'-A_q'}A_{q}'' \biggr|_{y=y_q}=0\ ,\\[1.5ex]
&f_{q+1}'-2A_{q+1}' f_{q+1}+\frac{f_q-f_{q+1}}{A_{q+1}'-A_q'}A_{q+1}'' \biggr|_{y=y_q}=0\ ,
\end{align}
with $q=1, \cdots, N-1$.

\section{The hermiticity condition}
\label{app:hermiticity}
We here derive the hermiticity condition presented in Eq.\,\eqref{eq:hermiticity}. Eq.\,\eqref{eq:schrodinger} has the form of
the Schr{\"o}dinger equation,
\begin{align}
\label{eq:schrodinger-like_app}
    \left[\frac{d^2}{dz^2}+H_s(z)\right]\tilde f_{s:n}=-m_n^2 \tilde f_{s:n}\ ,
\end{align}
where $H_s(z)$ is a function of $z$ and does not include a derivative with respect to $z$, the subscript $s$ denotes a function
for the subregion $s$ and $:n$ represents the eigenfunction with eigenvalue $m^2_n$. 
Let us now consider two modes $n=n_1,n_2$, and it is easy to find the following relation,
\begin{align}
\begin{split}
\label{Schdiff}
   &\sum_{s=1,2} \int_{\text{subregion $s$}} dz \left\{ \tilde f_{s:n_2} \left(\frac{d^2}{dz^2}+H_s(z)\right)\tilde f_{s:n_1}^*-\tilde f_{s:n_1}^*\left(\frac{d^2}{dz^2}+H_s(z)\right)\tilde f_{s:n_2}\right\}\\[1ex]
   &=(m_{n_2}^{2}-m_{n_1}^{*2}) \sum_{s=1,2} \int_{\text{subregion~$s$}} dz \, \tilde f_{s:n_1}^*\tilde f_{s:n_2} \, .
\end{split}
\end{align}
If the operator, $\frac{d^2}{dz^2}+H_s(z)$, is hermitian, the left-hand side of this relation must be zero.
By integration by parts, the left-hand side is reduced to
\begin{align}
\begin{split}
     &\sum_{s=1,2} \int_{\text{subregion~$s$}} dz\left\{ \frac{d}{dz}\left( \tilde f_{s:n_2} \frac{d}{dz}\tilde f^*_{s:n1}-\tilde f^*_{s:n_1}\frac{d}{dz}\tilde f_{s:n_2}\right)\right\}\\[1ex]
     &=
     \left[\tilde f_{1:n_2} \frac{d}{dz}\tilde f^*_{1:n1}-\tilde f^*_{1:n_1}\frac{d}{dz}\tilde f_{1:n_2}\right]_{z=z_{\rm UV}}^{z=z_{\rm I}}
     +\left[\tilde f_{2:n_2} \frac{d}{dz}\tilde f^*_{2:n_1}-\tilde f^*_{2:n_1}\frac{d}{dz}\tilde f_{2:n_2}\right]_{z=z_{\rm I}}^{z=z_{\rm IR}}\ ,
     \end{split}
\end{align}
where $z=z_{\rm UV,I,IR}$ denote the positions of the UV, intermediate and IR branes, respectively.
In the second line of the above equation, the square brackets represent $e.g.$
$[X(z)]_{z=z_{\rm UV}}^{z=z_{\rm I}} \equiv X(z_{\rm I})-X(z_{\rm UV})$ for $X(z)$ of a function of $z$.
Thus, the hermiticity condition is
\begin{align}
\label{eq:app_hermicity_condition}
    \left[\tilde f_{1:n_2} \frac{d}{dz}\tilde f^*_{1:n1}-\tilde f^*_{1:n_1}\frac{d}{dz}\tilde f_{1:n_2}\right]_{z=z_{\rm UV}}^{z=z_{\rm I}}
     +\left[\tilde f_{2:n_2} \frac{d}{dz}\tilde f^*_{2:n1}-\tilde f^*_{2:n_1}\frac{d}{dz}\tilde f_{2:n_2}\right]_{z=z_{\rm I}}^{z=z_{\rm IR}}=0\ .
\end{align}
Then, for $n_1=n_2$, the right-hand side of the relation \eqref{Schdiff} leads to
\begin{align}
    m_{n_1}=m_{n_1}^*\ ,
\end{align}
which indicates the real eigenvalues. For $m_{n_1}^*\neq m_{n_2}$, the hermiticity condition requires
\begin{align}
   \sum_s \int_{\text{subregion~$s$}} dz \tilde f_{s:n_1}^*\tilde f_{s:n_2}=0\ .
\end{align}
Thus, two eigenfunctions with different eigenvalues are orthogonal if the hermiticity condition \eqref{eq:app_hermicity_condition} is satisfied.
See also the main text for the hermiticity and orthogonality.

\bibliographystyle{utphys}
\bibliography{bib}

\end{document}